\documentclass[%
 reprint,
 amsmath,amssymb,
 aps,
pra,
superscriptaddress,
]{revtex4-2}

\usepackage{graphicx}% Include figure files
\usepackage{dcolumn}% Align table columns on decimal point
\usepackage{bm}% bold math
\usepackage{amsmath}
\usepackage{siunitx}

\usepackage{geometry}
\usepackage{longtable}
\usepackage{booktabs}

\usepackage{makecell} 
\usepackage{changes}

\usepackage{xr}
\begin{document}

\preprint{APS/123-QED}

\title{Large Transverse Thermoelectric Effect in Weyl Semimetal TaIrTe$_4$ Engineered for Photodetection}

\author{Morgan G. Blevins}
\affiliation{Department of Electrical Engineering and Computer Science, Massachusetts Institute of Technology, Cambridge, MA 02139, USA}
\affiliation{Draper Scholar, The Charles Stark Draper Laboratory, Inc., Cambridge, MA 02139, USA}
 % \email{mblevins@mit.edu}
\author{Xianglin Ji}
\affiliation{Department of Mechanical Engineering, Massachusetts Institute of Technology, Cambridge, MA 02139, USA}
\author{Vivian J. Santamaria-Garcia}
\affiliation{School of Engineering and Sciences, Tecnológico de Monterrey, Monterrey, 64700, Mexico.}
\author{Abhishek Mukherjee}
\affiliation{Department of Electrical Engineering and Computer Science, Massachusetts Institute of Technology, Cambridge, MA 02139, USA}
\author{Thanh Nguyen}
\affiliation{Department of Nuclear Science and Engineering, Massachusetts Institute of Technology, Cambridge, MA 02139, USA}
\author{Mingda Li}%
\affiliation{Department of Nuclear Science and Engineering, Massachusetts Institute of Technology, Cambridge, MA 02139, USA}
\author{Svetlana V. Boriskina}%
\affiliation{Department of Mechanical Engineering, Massachusetts Institute of Technology, Cambridge, MA 02139, USA}
\email{sborisk@mit.edu} %% email address is required

\date{\today}% It is always \today, today,

\begin{abstract}
Anomalous local photocurrent generation via second-order nonlinear and thermoelectric responses is a signature of many topological semimetals. The emergence of these photocurrents is inherently linked to symmetry breaking and anisotropy of their crystal lattices. Studies of type-II Weyl semimetals of group C$_{2v}$ (WTe$_2$, MoTe$_2$, TaIrTe$_4$) have reported anomalous, nonlocal photocurrents localized to crystals edges or far from electrodes, which are highly dependent on the geometry of the material sample. While originally attributed to a nonlinear charge current response, it was recently shown that these currents could instead be attributed to the anisotropic Seebeck coefficients of the materials. Here, we confirm that anomalous photocurrents observed in TaIrTe$_4$ under either visible or far-infrared far-field illumination originate from the large transverse thermoelectric effect. We engineer the mutual orientation of crystal edges and electrodes as well as the thermal environment of TaIrTe$_4$ to control and amplify its spatial photocurrent response. We show that substrate engineering can locally enhance photocurrent. This framework of thermal device engineering can enable broadband photo detection schemes by leveraging spectral and spatial dependence of photocurrents for applications like wavefront sensing, beam positioning, and edge detection.
\end{abstract}

\maketitle

% \section{Introduction}
Anomalous photocurrent generation in topological semimetals has recently attracted a lot of interest and sparked considerable controversy. The emergence of these photocurrents via the bulk photovoltaic effect (BPVE) is inherently linked to symmetry breaking and thus expected in the family of inversion symmetry (I) breaking Weyl semimetals (WSMs). I-breaking WSMs are especially interesting for infrared (IR) photo detection applications because their electronic band structures host low-energy optical transitions in the vicinity of the Weyl nodes, enabling strong IR absorption. In addition, the nontrivial topology of these materials gives rise to an enhanced BPVE, driven by large Berry curvature and associated nonlinear optical responses.

Shift currents driven by BPVE response under linearly polarized light and the currents originating due to circular photogalvanic effect (CPGE) were both experimentally measured in several WSM materials~\cite{osterhoudt_ColossalMidinfraredBulk_2019, ma_DirectOpticalDetection_2017}. 
However, WSMs are rich with other unique linear and nonlinear responses due to their topology and broken crystal symmetries. This impacts their optical absorption~\cite{tsurimaki_LargeNonreciprocalAbsorption_2020} as well as charge and heat transport properties~\cite{ pajovic_IntrinsicNonreciprocalReflection_2020,blevins_NonreciprocalHyperbolicSurface_2025a}, which should be taken into account when interpreting photocurrent response. Indeed, it was recently shown that anomalous photocurrents in the C$_{2v}$ crystal group of Type-II WSMs (WTe$_2$, MoTe$_2$, TaIrTe$_4$) could be attributed to an anisotropic photo-thermoelectric (PTE) response~\cite{wang_VisualizationBulkEdge_2023}, rather than BPVE as previously claimed. 

In this study, we use scanning photocurrent microscopy (SPCM) (Figure~\ref{fig:intro}A) and multi-physics modeling based on the Shockley-Ramo theory~\cite{song_ShockleyRamoTheoremLongrange_2014} to provide further evidence of the PTE origin of anomalous photocurrents in TaIrTe$_4$. We show that because TaIrTe$_4$ is a highly anisotropic p $\times$ n-type conductor~\cite{mutch_NTypeTransverseThermoelectrics_2022} with $n$- and $p$-type conductivities along its $a$ and $b$ crystal axes, it exhibits transverse PTE under specific device geometries and normal-incidence illumination. We further demonstrate that the transverse-PTE interpretation also extends to the long-wavelength IR photocurrent response of TaIrTe$_4$, where any strong anomalous BPVE stemming from the Weyl node topology would be expected to strongly contribute. We show how this response can be engineered, both to elucidate its physical origin and to selectively enhance or suppress local photocurrents for specific light detection schemes. 

\subsection{Anomalous photocurrents in $\text{TaIrTe}_4$}

\begin{figure*}
    \includegraphics[width=0.99\textwidth]{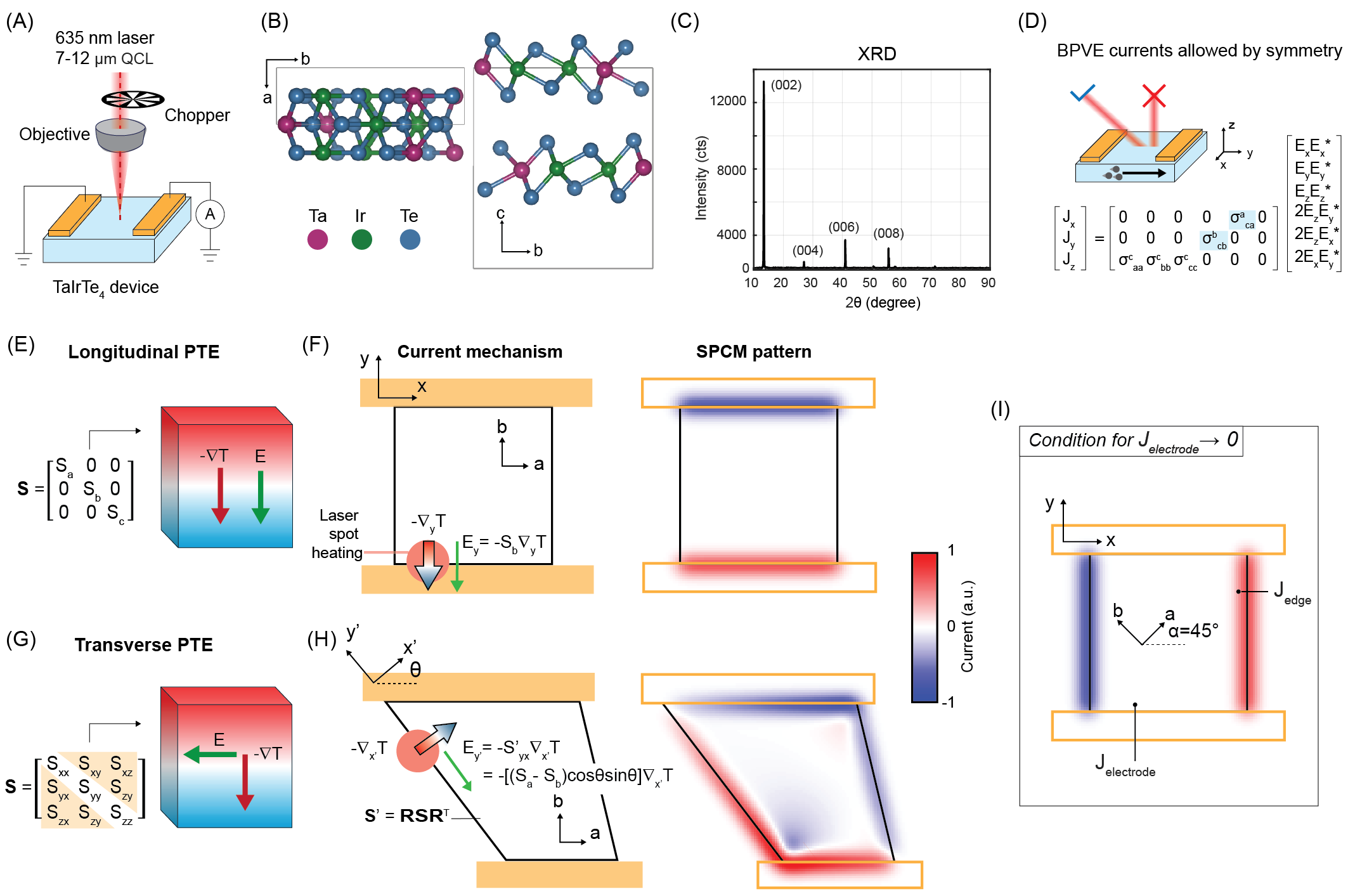}
    \caption{\textbf{Experimental overview and PTE mechanisms.} (A) Scanning photocurrent microscopy is used to characterize the photocurrent response of TaIrTe$_4$ devices. (B) TaIrTe$_4$ crystallizes in a layered, non-centrosymmetric orthorhombic crystal structure. (C) The XRD spectrum of our TaIrTe$_4$ crystal at ambient conditions shows prominent (00$\ell$) reflections, consistent with the Pmn2$_1$ space group. (D) Second-order NLO photocurrents are only allowed under non-normal incidence illumination according to the NLO tensor $\sigma^{(2)}_{ijk}$. (E) Under the longitudinal PTE, a temperature gradient drives a parallel electric field, (F) which can manifest in SPCM at the electrode-material interface due to partial obstruction of the laser spot, resulting in opposite-sign currents at either electrode. (G) In the transverse PTE, off-diagonal Seebeck tensor elements drive an electric field transverse to the temperature gradient direction. (H) In a p$\times$n-type conductor like TaIrTe$_4$, this can manifest as highly nonlocal edge currents on the off-axis edges. 
    (I) A device configuration for vanishing electrode current ($J_{electrode}$) with persistent edge currents ($J_{edge}$) for TaIrTe$_4$: a rectangular device with the a-axis at $\alpha=45^{\circ}$ relative to the electrodes. The SPCM maps in (F,H,I) are 2-D Shockley-Ramo simulations for TaIrTe$_4$ at $T=300$~\si{\kelvin} with $S_{a} = -6$ \si{\micro\volt\per\kelvin}, $S_{b} = 27$ \si{\micro\volt\per\kelvin}, $\sigma_a=4.91\times 10^5$ \si{\per\ohm\per\meter}, and $\sigma_b=1.1 \times 10^5 $ \si{\per\ohm\per\meter}. 
    }
    \label{fig:intro}
\end{figure*}

At ambient conditions, TaIrTe$_4$ crystallizes in a layered, non-centrosymmetric orthorhombic crystal structure with the space group Pmn2$_1$ and point group C$_{2v}$~\cite{lim_TemperatureinducedInversionSpinphotogalvanic_2018}. This structure lacks inversion symmetry but has a mirror plane in the b–c plane with a glide mirror symmetry along the b-axis, Fig~\ref{fig:intro}B ~\cite{jiang_ProbingInterplayTopological_2025}. The XRD spectrum of our TaIrTe$_4$ crystal was collected at ambient conditions, with Cu K$\alpha$ source radiation ($\lambda = 1.5406$ \si{\angstrom}) (Figure~\ref{fig:intro}C). We observe strong ($\ell$00) reflections indicating preferential growth along the [001] direction (c-axis), consistent with previous reports~\cite{tang_StrongcouplingAnisotropicSwave_2021}, confirming the Pmn2$_1$ space group.

The second-order BPVE response is characterized by the nonlinear second order conductivity tensor $\sigma^{(2)}_{ijk}$, resulting in the dc photocurrent generation $J_{i} = \sigma^{(2)}_{ijk}E_{j}E^*_{k}$, where $E_{j}$ is the ${j}-th$ component of the incident time-varying electric field (${j}=x,y,z$).
For TaIrTe$_4$, symmetry considerations dictate that the non-vanishing second order conductivity tensor elements are:
\begin{equation}
    \sigma_{aac}=\sigma_{aca}, \sigma_{bbc}=\sigma_{bcb}, \sigma_{caa}, \sigma_{cbb}, \sigma_{ccc}.
\end{equation} 
Accordingly, a field with a non-zero out-of-plane component $E_z$ is required to create in-plane (a-b plane) photocurrents (Figure~\ref{fig:intro}D)~\cite{lim_TemperatureinducedInversionSpinphotogalvanic_2018,shao_NonlinearNanoelectrodynamicsWeyl_2021}.

Despite the C$_{2v}$ symmetry forbidding 2nd-order nonlinear optical (NLO) photocurrents under normal-incidence illumination, multiple studies reported anomalous in-plane photocurrents in WTe$_2$ and TaIrTe$_4$ under such condition. These currents are typically localized to edges of the flakes and often extend far away from the Ohmic contacts~\cite{shao_NonlinearNanoelectrodynamicsWeyl_2021, wang_VisualizationBulkEdge_2023}. They were originally attributed to a 3rd order NLO charge current response~\cite{ma_DirectOpticalDetection_2017}. 

Subsequent work demonstrated that the use of a scanning metallic tip provides an out-of-plane field, $E_c$, and linked the measured currents to the shift current tensors $\sigma_{aac}$ and $\sigma_{bbc}$~\cite{shao_NonlinearNanoelectrodynamicsWeyl_2021}. In that study, the edge currents were attributed to the mirror symmetry breaking at the crystal edges, which was argued to permit new, nonzero shift current tensors. Modeling using $\sigma_{aaa}$ and $\sigma_{abb}$ values was shown to reproduce the experimental observations. 
% Space group is Pmn2$_1$.

However, it was recently revealed that these anomalous currents could instead originate from an anisotropic PTE response of materials with C$_{2v}$ symmetry~\cite{wang_VisualizationBulkEdge_2023}. The local photocurrent flow in WTe$_2$ and TaIrTe$_4$ was found to be consistent with a PTE current in a material with a highly anisotropic Seebeck coefficient. 

With this interpretation, the anomalous photocurrent response is decoupled from the BPVE mechanism, instead arising from the anisotropy of the in-plane crystal lattice. However, the PTE analysis of currents in TaIrTe$_4$ was limited to visible and mid-IR illumination conditions~\cite{deng_TuningDecayLength_2025}. In contrast, any anomalous contributions arising from electronic transitions near the Weyl nodes are expected to emerge under long-wavelength IR (LWIR) illumination ($\sim7-14$~\si{\micro\meter}), since the Weyl points and Fermi arcs in TaIrTe$_4$ lie 50–100 meV above the Fermi energy~\cite{xing_SurfaceSuperconductivityType_2020}. In this work, we investigate the photocurrent response of TaIrTe$_4$ in this LWIR regime and propose strategies to control it via thermal landscape engineering.

\subsection{Transverse PTE in $\text{TaIrTe}_4$ arising from its p$\times$n-type conductance}

NLO photocurrents are typically measured via the SPCM technique (Figure~\ref{fig:intro}A) on material samples with metal electrodes to enable electrical readout. In the measurement, a tightly focused laser beam is raster-scanned across the device, and the photocurrent is recorded without an external bias as a function of the beam position. The non-uniform illumination of a device can cause temperature gradients, $\nabla T$, and longitudinal electric fields, $E = \textbf{S}\cdot(-\nabla T)$, where $\textbf{S}$ is the Seebeck coefficient. These fields drive longitudinal PTE currents, $J = -\boldsymbol{\sigma} \textbf{S} \nabla T$, (where $\boldsymbol{\sigma}$ is the electrical conductivity tensor) often originate at the device-electrode interface, where the electrode truncates the beam and drives a large $\nabla T$ (Figure~\ref{fig:intro}E-F). The direction of this current is determined by the carrier type ($p$/$n$).

However, in an anisotropic material or a material under symmetry-breaking external stimuli, $\textbf{S}$ is a second-rank tensor. 
If this tensor has non-zero off-diagonal ${S_{ij}}$ elements, it can enable a transverse PTE, where an electric field is generated in the direction perpendicular to a temperature gradient (Figure~\ref{fig:intro}G): 
\begin{equation}
    E_i = S_{ij}(-\nabla_j T).
\end{equation}
Transverse PTE can only be observed under specific conditions, such as:
\begin{itemize}
    \item in a unipolar material under an external magnetic field, which induces a perpendicular electric field (known as the Nernst effect~\cite{goldsmid_IntroductionThermoelectricity_2016, behnia_NernstEffectMetals_2016})
    \item in materials with anisotropic transport properties~\cite{uchida_ThermoelectricsLongitudinalTransverse_2022,song_FieldGuideMaterials_2025}
\end{itemize}
In the Nernst effect, magnetic order or an applied magnetic field breaks time-reversal symmetry. This gives rise to antisymmetric off-diagonal components $S_{xy}$ in the thermoelectric tensor. Then, in the presence of an external dc magnetic field $\vec{B}$ or spontaneous magnetization $\vec{M}$, a transverse electric field emerges with $S_{xy} \neq S_{yx}$ proportional to $\vec{B}$ or $\vec{M}$. This effect has been found to be especially strong in topological semimetals~\cite{li_ColossalNernstPower_2022, han_QuantizedThermoelectricHall_2020, pan_UltrahighTransverseThermoelectric_2022}.

In materials with anisotropic thermoelectric tensors, transverse PTE arises from diagonal anisotropy combined with excitation geometry. Even if the tensor is diagonal in its principal crystal axes, 
\begin{equation} 
    \textbf{S}=
    \begin{pmatrix}
    S_{a} & 0 & 0\\
    0 & S_{b} & 0\\
    0 & 0 & S_{c}\\
    \end{pmatrix},
\end{equation}
when $\nabla T$ is applied at an angle $\theta$ to these axes, the rotated $\textbf{S}'$ tensor has off-diagonal elements: $S'_{xy}\propto (S_{a}-S_{b})\sin\theta\cos\theta$.
As a result, a transverse electric field is generated purely because of material anisotropy and axes misalignment, without breaking time-reversal symmetry (Figure~\ref{fig:intro}H). 

The anisotropic thermoelectric tensor can be engineered in superlattices~\cite{zhou_DrivingPerpendicularHeat_2013} and artificially tilted multilayers~\cite{ando_AdiabaticTransverseThermoelectric_2025}. It is also observed in natural p$\times$n-type conductors and goniopolar materials, i.e., materials that act as either $p$-type or $n$-type conductors depending on the crystallographic direction~\cite{ohsumi_TransverseThermoelectricConversion_2024, he_FermiSurfaceGeometrical_2019, manako_LargeTransverseThermoelectric_2024, pan_UltrahighTransverseThermoelectric_2022}.

TaIrTe$_4$ is a p$\times$n-type material, which exhibits opposite-sign anisotropic Seebeck coefficients along the a and b in-plane crystal axes, $S_{a}\neq S_{b}$, reported as $S_{a}\approx -6$ \si{\micro\volt\per\kelvin} and $S_{b}\approx 27$ \si{\micro\volt\per\kelvin} at room temperature ~\cite{mutch_NTypeTransverseThermoelectrics_2022}, with strong conductivity anisotropy ($\sigma_a=4.91\times 10^5$ \si{\per\ohm\per\meter}, $\sigma_b=1.1 \times 10^5 $ \si{\per\ohm\per\meter}~\cite{zhang_RoomTemperatureFieldfree_2023}). 
The origin of the p$\times$n TE response in TaIrTe$_4$ is the coexistence of anisotropic electron and hole pockets close to the material’s Fermi level~\cite{khim_MagnetotransportHaasvanAlphen_2016, koepernik_TaIrTe4TernaryTypeII_2016, soluyanov_TypeIIWeylSemimetals_2015}. 
These pockets enable opposite thermopower contributions along orthogonal crystal axes. 
% This behavior is not tied to the topology of the TaIrTe$_4$ electronic band structure~\cite{mutch_NTypeTransverseThermoelectrics_2022}.
Thus, when a $\nabla T$ is applied to TaIrTe$_4$ at an angle $\theta$ to the a-b crystal axes, a transverse PTE is activated. 

For specific geometric configurations of thermally anisotropic materials --- characterized by an angle $\alpha$ between the crystal a-axis and electrode, and an angle $\theta$ between the crystal edge and electrode --- edge currents can persist even when currents at the electrode interface vanish. 
An example geometry for a rectangular device ($\theta=0^{\circ}$) is shown for TaIrTe$_4$ in Figure~\ref{fig:intro}I. 
In this case $J_{electrode}= -\sigma_{yy} S_{yy} \nabla_y T$ and $J_{edge} = -\sigma_{yx} S_{yx} \nabla_x T$. To maximize edge current while minimizing electrode current we need to maximize the ratio 
\begin{equation}
    \left|\frac{\sigma_{yx} S_{yx}}{\sigma_{yy} S_{yy}}\right|=
    \left|
    \frac{\sigma_{yx}(S_{a}-S_{b})\sin\alpha\cos\alpha}
        {\sigma_{yy}(S_{a}\sin^2\alpha+S_{b}\cos^2\alpha)}
    \right|.
\end{equation}
This condition is met for $\alpha=45$\si{\degree}, which results in minimal current at the electrodes while edge currents persist~\cite{wang_VisualizationBulkEdge_2023}.

The figure of merit of the transverse thermoelectric effect is $z_{xy}T$ ~\cite{ goldsmid_ApplicationTransverseThermoelectric_2011}:
\begin{equation}
    z_{xy}T = \frac{S_{xy}^2}{\rho_{xx}\kappa_{yy}}T,
\end{equation}
where $\rho_{xx}$ and $\kappa_{yy}$ are the isothermal electrical resistivity and thermal conductivity respectively. 
Based on the parameters of Ref.~\cite{mutch_NTypeTransverseThermoelectrics_2022}, TaIrTe$_4$ has a $z_{xy}T\approx1.5\times10^{-3}$ at room temperature, which is comparable to the large thermoelectric performance recently identified in the conductor LaPt$_2$B ~\cite{manako_LargeTransverseThermoelectric_2024}. This large transverse PTE FoM without a magnetic bias is notable. While the thermally driven currents were previously measured and justified, here we place the PTE in TaIrTe$_4$ in context of the state of the art in transverse thermoelectric devices. 

\begin{figure*}
    \includegraphics[width=0.99\textwidth]{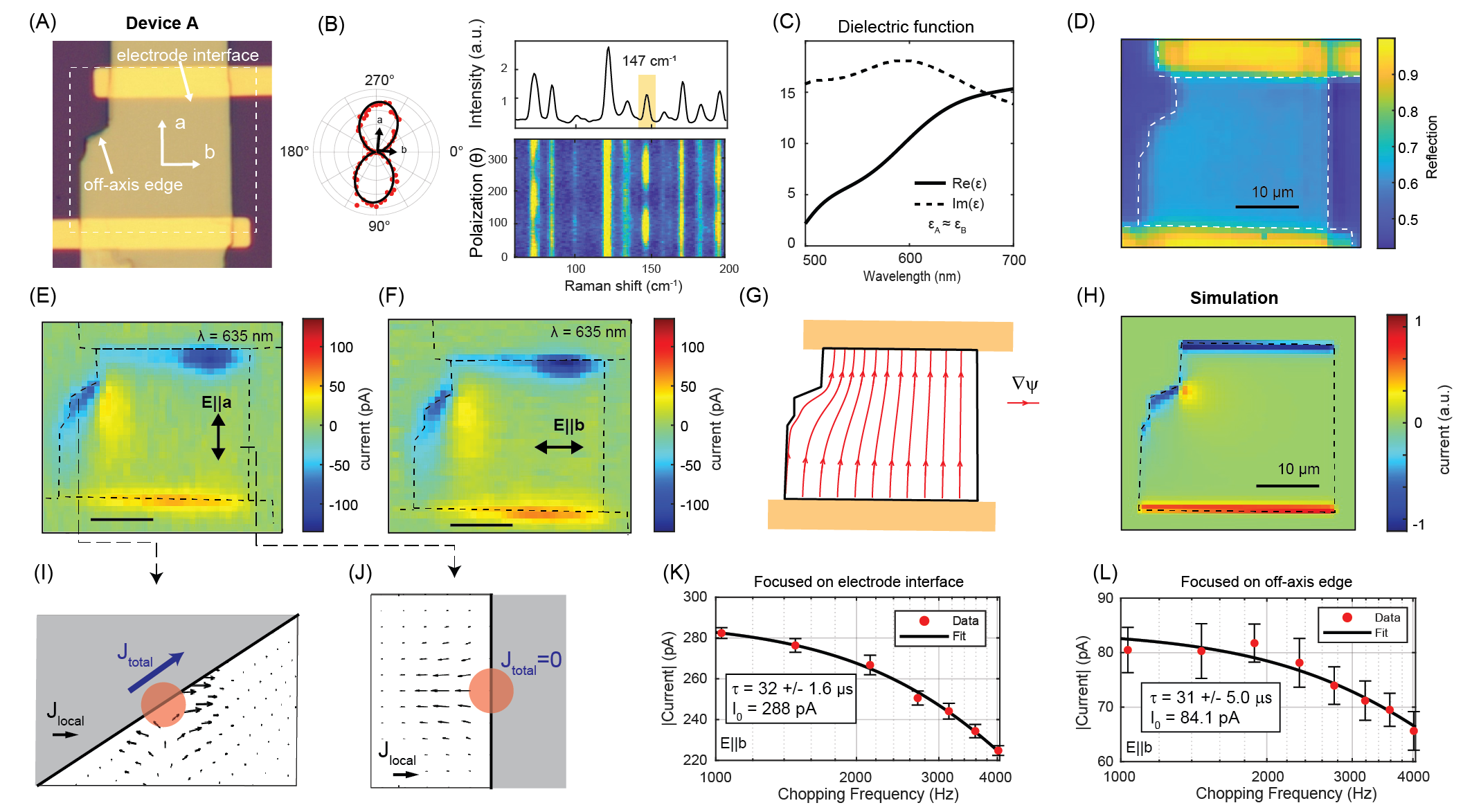}
\caption{
    \textbf{Longitudinal and transverse PTE in a TaIrTe$_4$ device at 635 nm illumination.} (A) A single-crystal TaIrTe$_4$ photodetector with the a-axis aligned along the natural device edge on the right, as identified by (B) angle-resolved polarized Raman spectroscopy. The 147~\si{\per\centi\meter} peak has 180\si{\degree} periodicity with maximum intensity along the a-axis. (C) The experimentally measured dielectric function $\varepsilon_{a,b}(\omega)$ of TaIrTe$_4$ in the visible and near-infrared region, approximated as isotropic in-plane. SPCM maps of (D) reflection and photocurrent were measured with 17~\si{\micro\watt} linearly polarized 635 \si{\nano\meter} light for (E) $E\parallel a$ and (F) $E\parallel b$, showing no polarization dependence. (G) The weighting field $\nabla\psi$ of the device, as visualized with streamlines, is taken into account along with the anisotropic Seebeck tensor of TaIrTe$_4$ in (H) the simulated photocurrent pattern using Shockley-Ramo theory, which is in good agreement with experiments for both $E\parallel a$ and $E\parallel b$. (I) We simulate the local photocurrent vector field $\vec{J}_{\text{loc}}(\mathbf{r})$ at the off-axis crystal edge, showing that net photocurrent is measured between the electrodes due to a nonzero transverse PTE current component parallel to the weighting field: $\vec{J}_{\text{loc}}(\mathbf{r}) \parallel \nabla\psi$. (J) At the a-axis edge there is zero net current because $\vec{J}_{\text{loc}}(\mathbf{r}) \perp \nabla\psi$ in that region. The chopper frequency roll-off was measured to extract the photocurrent time response at the (K) electrode interface, $\tau = 32 \pm 1.6$ \si{\micro\second}, and (L) the off-axis edge, $\tau = 31 \pm 5.0$ \si{\micro\second}.}
    \label{fig:635nm-PTE}
\end{figure*}

\begin{figure*}
    \includegraphics[width=0.99\textwidth]{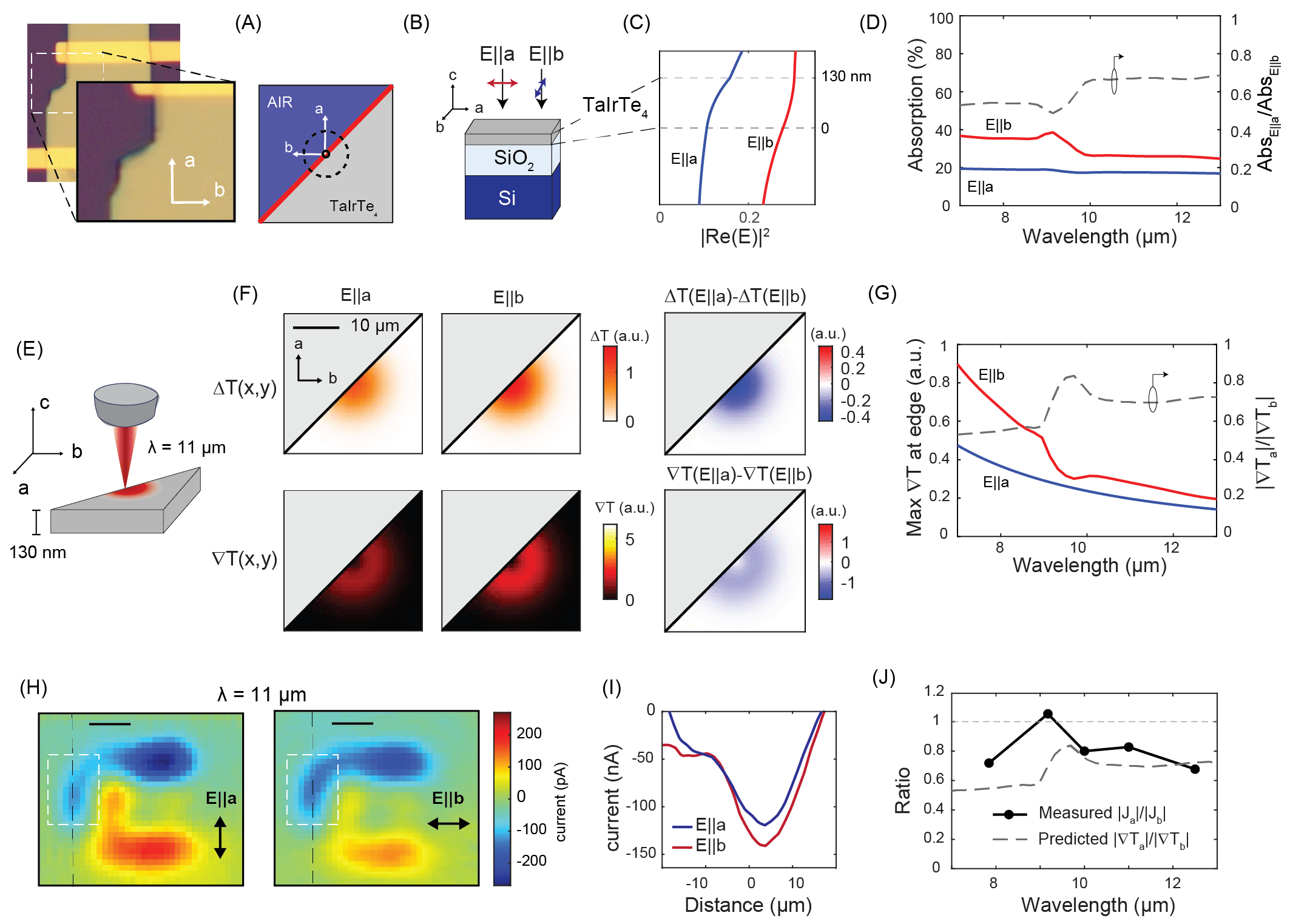}
    \caption{
    \textbf{Long-wave infrared photocurrent response of TaIrTe$_4$.}
    The LWIR response is investigated at (A) the off-axis edge of device A, which is (B) 130~\si{\nano\meter} thick on a SiO$_2$(285~nm)/Si substrate. The dielectric functions of TaIrTe$_4$ in the LWIR is metallic along the a-axis and dielectric along the b-axis, which is apparent in (C) the calculate E-field intensity in the TaIrTe$_4$ flake for $E \parallel a$ versus $E \parallel b$ (shown for $\lambda=11$~\si{\micro\meter}. (D) The modeled absorption into the flake is higher for $E \parallel b$. (E) The thermal response in TaIrTe$_4$ for focused laser illumination in the LWIR is highly dependent on light polarization due to the in-plane optical anisotropy. (F) This is apparent in the temperature $T$ and temperature gradient $\nabla T$ profiles of TaIrTe$_4$ for a focused laser spot, in-plane and out-of-plane. (G) Accounting for the wavelength-dependent optical absorption, the calculated maximum $\nabla T_x$ at the edge is larger for $E\parallel b$. (H) LWIR SPCM photocurrent maps were measured for five wavelengths between 7.7-12~\si{\micro\meter} for $E||a$ and $E||a$. The results for 11~\si{\micro\meter} are shown for 285~\si{\micro\watt} laser power. The black dashed lines correspond to (I) the plotted profile of the edge-current, showing $|I^{edge}_b| >|I^{edge}_a|$. (J) The spectral dependence of the the off-axis edge photocurrent matches the trend predicted by our $\nabla T(\omega)$ calculation.} 
    \label{fig:IR-results}
\end{figure*}

\begin{figure*}
    \includegraphics[width=0.99\textwidth]{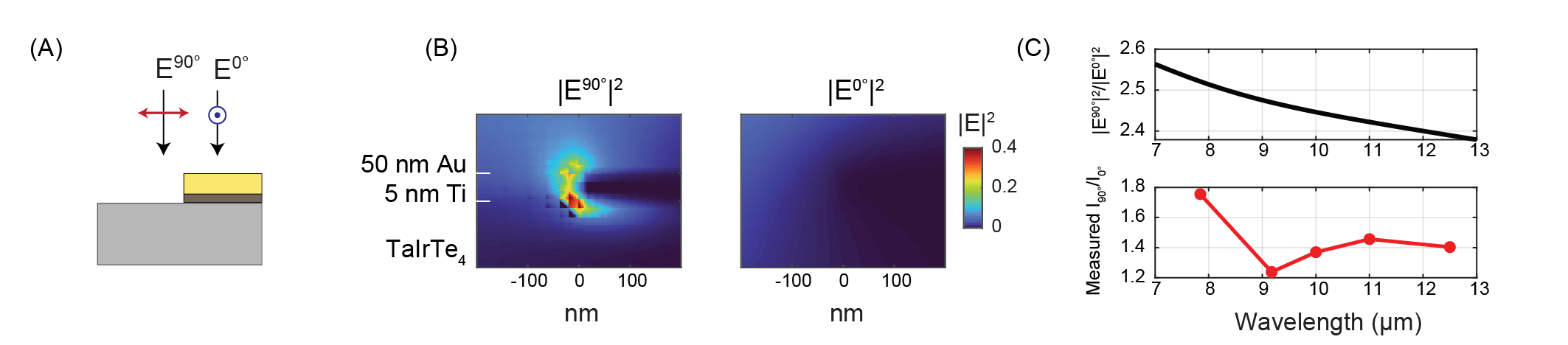}
    \caption{
    \textbf{Plasmonic impact of gold electrodes in LWIR.}
    (A) There is a well known plasmonic enhancement when the incident light is polarized perpendicular to the electrode interface, shown by (B) FDTD calculated E-field intensity for $E^{90^{\circ}}$ versus $E^{0^{\circ}}$. (C) In the 7-13~\si{\micro\meter} spectral region, the enhancement is calculated as $\sim2.5\times$. This is compared to the measured trend of the maximum $I_{90^{\circ}}/I_{0^{\circ}}$ at the bottom electrode interface of Device A (full data in Figure~\ref{fig:dev-A-IR}).} 
    \label{fig:plasmonic}
\end{figure*}

\subsection{Thermally driven photocurrents at 635 nm illumination}
Two devices (labeled A and B) were fabricated via mechanical exfoliation and transfer of TaIrTe$_4$ flakes onto 300-\si{\nano\meter}-SiO$_2$/Si substrates, followed by deposition of gold (Au) electrodes for current collection. The I-V curve for each device is linear, confirming good Ohmic contact (Figure~\ref{fig:IV-curve}). In device A (Figure~\ref{fig:635nm-PTE}A), which is 130~\si{\nano\meter} thick, the crystal $a$-axis is aligned along the natural device edge on the right, while an off-axis crystal edge is present in the left corner. The crystal axis orientation was determined using angle-resolved polarized Raman spectroscopy~\cite{liu_RamanSignaturesBroken_2018} (Figure~\ref{fig:635nm-PTE}B).
The optical response of TaIrTe$_4$ in the visible range was measured with an imaging ellipsometer, from which the complex dielectric function was extracted (Figure~\ref{fig:635nm-PTE}C). The response is approximately isotropic in-plane ($\varepsilon_A\approx\varepsilon_B$) in this spectral region, consistent with prior measurements~\cite{shao_NonlinearNanoelectrodynamicsWeyl_2021}.

The SPCM reflection and photocurrent response were measured under normal illumination using 635 \si{\nano\meter}-wavelength linearly polarized light ($E\parallel a$ and $E\parallel b$) with 17~\si{\micro\watt} power and no external bias (Figure~\ref{fig:635nm-PTE}D–F).

Normal incident light has no out-of-plane component necessary to generate an NLO current via the BPVE. Nevertheless, the experimental photocurrent maps reveal a response along the off-axis crystal edge of the flake extending far from the Ohmic contacts (Figure~\ref{fig:635nm-PTE}E–F). These photocurrents are insensitive to the 635~\si{\nano\meter} light polarization direction ($E\parallel a$ versus $E\parallel b$), a hallmark of PTE currents in materials with isotropic absorption (confirmed by polarization-insensitive reflectance maps; Figure~\ref{fig:dev-A-635nm}).

To confirm the thermoelectric origin of these currents and explain the spatial pattern in the device, we used Shockley-Ramo theory to calculate the local photocurrent transport originating from the anisotropic $\textbf{S}$ of TaIrTe$_4$~\cite{song_ShockleyRamoTheoremLongrange_2014}. The current recorded at remote contacts is obtained by the Shockley–Ramo collection integral,
\begin{equation}
\label{eq:current-total}
I_{\rm tot}=\iint_{\mathbf{r}} \mathbf{J}_{\rm loc}(\mathbf{r})\cdot\nabla\psi(\mathbf{r})\,\mathrm{d}^2\mathbf{r},
\end{equation}

\noindent where $\psi(\mathbf{r})$ is the device-dependent weighting potential, visualized in Figure~\ref{fig:635nm-PTE}G, and $\mathbf{J}_{\rm loc}$ is the local photocurrent vector field when the laser is focused at location $\mathbf{r}$. Equation~\eqref{eq:current-total} captures the long-range, geometry-sensitive collection of local currents measured in SPCM and is agnostic to the microscopic photocurrent generation mechanism. To evaluate local photocurrents driven by PTE sources we apply $\mathbf{J}_{\rm loc}=-\boldsymbol{\sigma}\,\mathbf{S}\nabla T$. To calculate $T(x,y,z)$, we solve the 3D anisotropic heat equation in the flake accounting for the laser spot heating and material parameters including thermal conductivity $\kappa_{a,b,c}$, density $\rho$, heat capacity $C_p$, and thermal boundary conductance $G$~\cite{zhu_AnisotropicHotCarrier_2025,zhang_RoomTemperatureFieldfree_2023,liu_FirstprinciplesStudyLattice_2016,hunter_InterfacialThermalConductance_2020}.
The example SPCM maps in Fig~\ref{fig:intro}F,H,I are simulated for a simplified 2-D case, accounting for $T(x,y)$. Note~\ref{sec:SPCM-sim} contains the full methodology and Table~\ref{tab:parameters} contains the material parameters used. We justify the use of the Seebeck coefficients from literature in Note~\ref{sec:Seebeck-scan} (Figure~\ref{fig:seebeck_maps} and~\ref{fig:seebeck_cond_sensitivity}).

The resulting simulations (Figs.~\ref{fig:635nm-PTE}H) are in good agreement with the experimental SPCM maps and can be understood by examining the relationship between the local current and the weighting field. At the off-axis edge, the $\nabla T$ generates a transverse PTE current that has a component parallel to the weighting field ($\vec{J}_{\text{loc}} \parallel \nabla\psi$), enabling current collection (Figure~\ref{fig:635nm-PTE}I). Along the natural $a$-axis edge, however, the local photocurrent is oriented perpendicular to the weighting field gradient ($\vec{J}_{\text{loc}} \perp \nabla\psi$), so no net current is collected at these edges (Figure~\ref{fig:635nm-PTE}J). These observations confirm that the dominant photocurrent mechanism is a transverse PTE effect. 

Furthermore, the device exhibits signal roll-off with increasing chopping frequency at the edge and electrode interface (Figure~\ref{fig:635nm-PTE}K-L) indicating that the temperature gradient cannot be maintained when the modulation speed exceeds the thermal diffusion time of the material. The large time responses measured ($\tau = 32 \pm 1.6$ \si{\micro\second} at the electrode interface and $\tau = 31 \pm 5.0$ \si{\micro\second} at the off-axis edge) provide further evidence for the PTE rather than nonlinear photocurrent mechanism. 

\subsection{Photocurrent response in the infrared}

We next investigated whether the LWIR photocurrent in TaIrTe$_4$ originates from the photothermal mechanisms (PTE) or from nonlinear optical effects (BPVE). To distinguish between these mechanisms, we probed the off-axis crystal edge of device A (Figure~\ref{fig:IR-results}A), where the photocurrent is expected to arise purely from PTE. This location avoids the electrode interfaces, where plasmonic enhancement from the Au contacts can also contribute to the photoresponse.

In the LWIR spectral region, the optical response of thin TaIrTe$_4$ flakes (10–100s~\si{\nano\meter}) depends strongly on wavelength, polarization, and flake thickness. This behavior arises from the highly anisotropic in-plane permittivity: along the a-axis, TaIrTe$_4$ is metallic and lossy ($\text{Re}[\varepsilon_a] < 0$, large $\text{Im}[\varepsilon_a]$), while along the b-axis it is dielectric ($\text{Re}[\varepsilon_b] > 0$, smaller $\text{Im}[\varepsilon_b]$) (Figure~\ref{fig:dielectric-func})~\cite{shao_NonlinearNanoelectrodynamicsWeyl_2021}. 
For the case of device A (thickness $t=130$~\si{\nano\meter}), we calculated the polarization dependent E-field intensity and absorption ($Abs$) in the flake via rigorous coupled-wave analysis (RCWA), finding that $Abs(E\parallel b)>Abs(E\parallel a)$ in the LWIR, Fig~\ref{fig:IR-results}B-D. The absorption peak near 9~\si{\micro\meter} for $E\parallel b$ is due to the resonant phonon vibration modes of SiO$_2$ between 9-9.5~\si{\micro\meter}~\cite{kitamura_OpticalConstantsSilica_2007}. This is not present for $E\parallel a$ because the material is highly metallic for that polarization state and light does not penetrate to the SiO$_2$ layer. 

To predict the expected PTE photocurrent, we calculated the temperature rise over ambient ($\Delta T(x,y)$) and thermal gradient ($\nabla T(x,y)$) in device A when the laser is focused near a crystal edge 45\si{\degree} to the a-axis. We solved the 2D heat equation, treating the laser spot as a Gaussian heat source and accounting for the insulating crystal edge-air boundary. In Figure ~\ref{fig:IR-results}F we show the case of $\lambda=11$~\si{\micro\meter}. Both $\Delta T$ and $\nabla T$ are larger for the chase of $E\parallel b$. Indeed, the calculated trend of $\nabla T$ as a function of laser wavelength (Figure~\ref{fig:IR-results}G) predicts that $\nabla T(E\parallel b) > \nabla T(E\parallel a)$ across the LWIR spectral region, thus predicting the same trend for $|I^{edge}_a|/|I^{edge}_b|$.

We experimentally measured polarization- and wavelength-dependent photocurrent maps under 7-12~\si{\micro\meter} illumination, ($\lambda=11$~\si{\micro\meter} in Figure~\ref{fig:IR-results}H, others in Figure~\ref{fig:dev-A-IR}). At the off-axis crystal edge, the photocurrent magnitude is larger for $E\parallel b$ than $E\parallel a$ (Figure~\ref{fig:IR-results}I-J). Plotting the spectral ratio of $|I^{edge}_a|/|I^{edge}_b|$ we see good agreement with the expected polarization trend of $\nabla T$. This agreement indicates that the wavelength- and polarization-dependent photocurrent originates from the linear optical anisotropy of TaIrTe$_4$ and is fully explained by the PTE mechanism, with no evidence of anomalous contributions from nonlinear effects such as the BPVE.

Notably, the trend at the TaIrTe$_4$-electrode contacts (50-nm Au on 5-nm Ti) follows the opposite polarization dependence: here, $|I_a|>|I_b|$. This is explained by well-known plasmonic enhancement effects, which are strongest when the incident polarization is perpendicular to the electrode ($E^{90^{\circ}}$, corresponding to $E\parallel a$ in our geometry) illustrated in Figure~\ref{fig:plasmonic}A~\cite{hong_PlasmonicHotElectron_2015, tielrooij_HotcarrierPhotocurrentEffects_2015}. This enhanced electric field confined to the TaIrTe$_4$-electrode interface maximizes absorption for $E^{90^{\circ}}$ and alters the temperature gradient at the interface, enhancing PTE current~\cite{echtermeyer_StrongPlasmonicEnhancement_2011a}. We calculated the E-field intensity enhancement at the Au/Ti electrode interface for $E^{90^{\circ}}$ and $E^{0^{\circ}}$ using FDTD (Figure~\ref{fig:plasmonic}B). The predicted wavelength dependence of $|E^{90^{\circ}}|^2/|E^{0^{\circ}}|^2$ shows an $\sim2.5\times$ enhancement, which generally agrees with the trend measured at the TaIrTe$_4$-electrode contact of device A (Figure~\ref{fig:plasmonic}C). 

Finally, we note that prior SPCM studies of TaIrTe$_4$ have used flakes with thickness of roughly $t\leq60$~\si{\nano\meter}~\cite{ma_DirectOpticalDetection_2017, lai_BroadbandAnisotropicPhotoresponse_2018,deng_TuningDecayLength_2025}. Our RCWA modeling reveals that for this low thickness, the absorption trend reverses: $Abs(E\parallel a)>Abs(E\parallel b)$, because the response becomes dominated by $\text{Im}[\varepsilon_A]$ (Figure~\ref{fig:IR-sim-30nm}). Consistently, prior studies on thin TaIrTe$_4$ report larger photocurrent for $E \parallel a$ than $E \parallel a$, in agreement with our analysis. This thickness-dependent crossover in polarization response further supports the PTE origin of the photocurrent and reconciles the apparently contradictory results across the literature.

\begin{figure*}
    \includegraphics[width=0.99\textwidth]{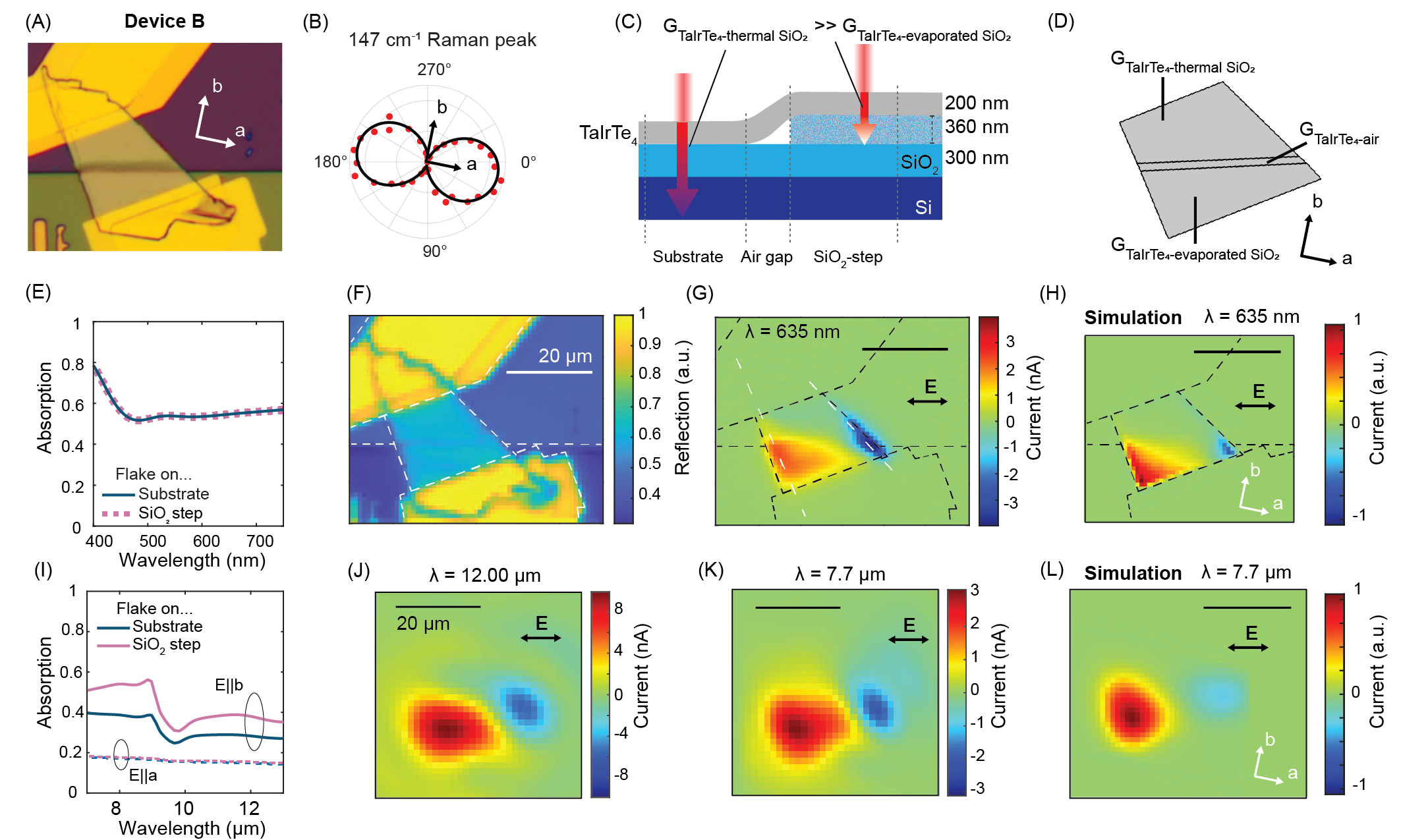}
    \caption{
    \textbf{Thermally engineered photocurrent response of TaIrTe$_4$.} (A) A 200 \si{\nano\meter} thick TaIrTe$_4$ flake was placed on a substrate, half suspended on top of a 360 \si{\nano\meter} evaporated SiO$_2$ step. (B) The crystal axes are identified by angle-resolved polarized Raman spectroscopy. (C) The reduced TBC $G_{\text{TaIrTe}_{4}\text{-SiO}_{2}}$ of the evaporated SiO$_2$ is illustrated with a shorter arrow compared to that of the thermally grown SiO$_2$. (D) The device consequently has three regions with different TBC: TaIrTe$_4$ on thermally grown SiO$_2$, a transition region where TaIrTe$_4$ is suspended over air, and TaIrTe$_4$ on evaporated SiO$_2$. (E) The simulated light absorption into the flake is the same on and off the SiO$_2$ step based on the measured $\varepsilon(\omega)$ in the visible. The device was measured with SPCM and the (F) reflection and (G) photocurrent maps at 635 \si{\nano\meter} illumination (19~\si{\micro\watt} power) show localization of photocurrent to the device edges in the SiO$_2$ step region of the flake. Accounting for the decreased TBC on the step, (H) the Shockley-Ramo simulated response is in good agreement with the measured response. (I) We predict a slight absorption enhancement for $E \parallel b$, and none for $E \parallel a$ on the SiO$_2$ step in the LWIR. SPCM with (J) $\lambda=12$~\si{\micro\meter} and (K) $\lambda=7.7$~\si{\micro\meter} (71 and 230 ~\si{\micro\watt} power respectively) for $\angle (\vec{E} , \vec{a})=10^{\circ}$ follow the same photocurrent pattern as the $635$~\si{\nano\meter} case and (L) the simulated photocurrent pattern accounting for the IR optical response replicates the observed pattern.}
    \label{fig:sio2-step-635nm-results}
\end{figure*}

\subsection{Thermally engineered photocurrent}

To further probe and control the PTE mechanism of the TaIrTe$_4$ photocurrent response, we engineered the thermal environment of another flake (Device B). A TaIrTe$_4$ flake was placed on a 300-\si{\nano\meter}-SiO$_2$/Si substrate and partially suspended over a 320~\si{\nano\meter} e-beam evaporated SiO$_2$ step, creating three regions with distinct thermal boundary conductance (TBC) $G$: TaIrTe$_4$ on thermally grown SiO$_2$, a narrow air-suspended transition, and TaIrTe$_4$ on evaporated SiO$_2$ (Figure~\ref{fig:sio2-step-635nm-results}A-C; Figure~\ref{fig:afm}). 
Because e-beam evaporated SiO$_2$ is more porous and amorphous than thermally grown oxide, we expect $G_{\text{TaIrTe}_4\text{-thermal SiO}_2} \gg G_{\text{TaIrTe}_4\text{-evaporated SiO}_2}$, with the air gap presenting the lowest TBC (Figure~\ref{fig:sio2-step-635nm-results}D; Figure~\ref{fig:thermal_contacts})~\cite{hopkins_EffectsSurfaceRoughness_2010}. Raman characterization confirmed this thermal confinement (Figure~\ref{fig:raman-spectrum-burn}). RCWA shows that the SiO$_2$ step does not affect optical absorption in the flake for visible wavelengths (Figure~\ref{fig:sio2-step-635nm-results}E), so photonic enhancements are not expected.

We measured device B under linearly polarized 635~\si{\nano\meter} illumination (Figure~\ref{fig:sio2-step-635nm-results}F-G). The off-axis crystal edge photocurrent was strongly enhanced in the SiO$_2$ step region on both flake edges, while the reflection map confirmed uniform local absorption across the step. Our modeling and characterization predicts that this enhancement arises from reduced heat dissipation: lower TBC in the gap and across the evaporated SiO$_2$ step area produces greater local heating and thus stronger PTE response. Our simulations reproduce this behavior when incorporating a two-order-of-magnitude reduction in TBC at the step and a minimal TBC for the air-suspended transition (Figure~\ref{fig:sio2-step-635nm-results}H). Such TBC variations are consistent with previous reports on mechanically transferred 2D materials, where interfacial quality can reduce TBC by up to two orders of magnitude~\cite{frausto-avila_ThermalBoundaryConductance_2022}. We note that the agreement between experiment and simulation slightly diverges for the right-side edge (negative current region): in experiment, the strong edge current extends slightly further below the SiO$_2$ step. We attribute this to a larger air gap between the flake and the substrate (as illustrated in Figure~\ref{fig:sio2-step-635nm-results}C) on the right-side edge. This is evident in the AFM topography maps of the device, where the air gap is measured as roughly twice as long on the right edge (Figure~\ref{fig:afm}B). Further, there are regions on the right-side edge below the step where the flake bends up by as much as 200~\si{\nano\meter} above the substrate, which further introduces air gaps which trap heat and enhance edge current.

We next measured SPCM maps under the LWIR laser illumination (Figure~\ref{fig:sio2-step-635nm-results}I-L). Absorption is predicted to increase by roughly $\sim 50\%$ for $E\parallel b$, but not for $E\parallel a$. We conducted SPCM measurements for $\angle (\vec{E} , \vec{a})=10^{\circ}$, and the resulting photocurrent map follows the same qualitative pattern as was observed under visible light illumination (Figure~\ref{fig:sio2-step-635nm-results}J-K), as expected for the PTE. We simulated the response in the LWIR spectral range, accounting for the anisotropic optical and thermal response, and found it to be in good agreement with experiment (Figure~\ref{fig:sio2-step-635nm-results}L).

The introduction of the SiO$_2$ step was also intended to induce strain in the crystal to modify the Seebeck tensor anisotropy. However, we used Raman spectroscopy to probe crystal lattice strain in a step-suspended flake (Figure~\ref{fig:raman-line}) and saw minimal evidence of TaIrTe$_4$ Raman peak shifting on the step~\cite{yang_RamanactiveModes1TWTe2_2019}. 

\subsection{Discussion}

We have shown that the PTE mechanism of photocurrent generation in TaIrTe$_4$ extends to the LWIR spectral range, and that engineering the thermal environment of this material allows for control and enhancement of the PTE photocurrents. This study reveals that careful consideration of the thermal environment of self-biased photodetectors utilizing Weyl semimetals like TaIrTe$_4$ is necessary to properly interpret the mechanisms of photocurrent generation and to avoid misidentification of anomalous or nonlinear photocurrents. It also highlights the opportunities for the performance improvement of PTE-based detectors via thermal environment engineering.

The shape of the TaIrTe$_4$ device edges and the asymmetry of device contacts can be engineered to control both the longitudinal and transverse PTE response in tandem and create complex local photocurrent response patterns. This position-dependent photoresponse could provide unique opportunities for beam positioning or edge detection. Recent work has explored how designing the relative angle between electrodes and the crystal axes can tune carrier decay length; this approach could be used in tandem with thermal engineering \cite{deng_TuningDecayLength_2025}.

Given the exceptional transverse thermoelectric figure of merit $z_{xy}T$ without magnetic bias, applications beyond photodetection should be pursued in TaIrTe$_4$, such as thermal energy recovery and cooling. Other materials exhibiting the anisotropic Seebeck coefficients and thermal conductivity can also provide useful platforms for PTE-based photodetection and energy recovery~\cite{manako_LargeTransverseThermoelectric_2024,uchida_ThermoelectricsLongitudinalTransverse_2022,ohsumi_TransverseThermoelectricConversion_2024}. The transverse thermoelectric effect is especially useful for enabling flat device geometries where thermoelectric conversion is needed in small areas for microelectronics. Making use of the transverse PTE in TaIrTe$_4$ and other materials could enable miniaturized thermal management systems that address waste heat through both radiative and conductive mechanisms.

The presence of similar (but more balanced) electron and hole pockets in a sister Weyl semimetal Td-WTe$_2$ was shown to yield an ultrahigh Nernst power factor under magnetic bias~\cite{pan_UltrahighTransverseThermoelectric_2022, zhao_AnisotropicMagnetotransportExotic_2015, zhu_QuantumOscillationsThermoelectric_2015}. The electronic band structure and Seebeck coefficients of TaIrTe$_4$ and other materials can be similarly modified by external stimuli, including magnetic field, strain, pressure, chemical doping, and temperature. These stimuli can trigger Lifshitz transitions resulting in either creation or annihilation of Fermi surface pockets, and the accompanying changes in the electronic and transport properties of materials~\cite{li_UltrahighSeebeckCoefficient_2024, luo_NonlinearInfraredPhotocurrent_2026, biesner_SpectroscopicTraceLifshitz_2021}. Stimuli-tuned enhancement of the PTE can unlock ultrahigh thermopower, which could be promising for low-temperature PTE energy harvesting devices ~\cite{han_QuantizedThermoelectricHall_2020}.

\section{Methods}

\subsection{Device fabrication}
Single crystals of TaIrTe$_4$ were synthesized using a solid-state solution method using Te flux. The TaIrTe$_4$ flakes were mechanically exfoliated from the bulk single crystal onto 300-\si{\nano\meter}-SiO$_2$/Si wafers with blue ProTapes Nitto SPV224 tape. Electrodes of 50-nm Au on 5-nm Ti were deposited onto identified flakes using the Heidelberg MLA 150 for photolithography and the AJA Model ATC metal evaporator for deposition.

To engineer device B, 360 \si{\nano\meter} of SiO$_2$ was evaporated onto the 300-\si{\nano\meter}-SiO$_2$/Si substrate dies using standard photolithography techniques and the Temescal FC-2800 for deposition. The TaIrTe$_4$ flakes were mechanically exfoliated onto the dies and flakes adhered to the SiO$_2$ gratings were identified afterward for electrode deposition. 

\subsection{Device characterization}
Device thickness was measure with an Asylum Instruments/Oxford Jupiter XR AFM (Note~\ref{fig:afm}; Table~\ref{tab:device-thicknesses}). The WITec Alpha300R Raman microscope with a 632.808~\si{\nano\meter} laser source was used to collect angle-resolved polarized Raman spectroscopy sweeps to identify the crystal axes of the TaIrTe$_4$ flakes.
The Accurion EP4 imaging ellipsometer was used to measure the dielectric permittivity of TaIrTe$_4$ in the visible and near infrared spectral bands. 

\subsection{Scanning Photocurrent microscope}
A scanning photocurrent microscope with a 40x, 0.4 NA reflective objective was used with a 635 \si{\nano\meter} laser diode ($\sim1$~\si{\micro\meter} diffraction limited spot size) and a 7-13 \si{\micro\meter} Block LaserTune quantum cascade laser (QCL) source ($\sim9-16$~\si{\micro\meter} diffraction limited spot size). Visible and infrared linear polarizers were placed in the beam path to control polarization. The sample was rotated in relation to the SPCM to control the linear polarization angle. An optical chopper was used at 3675 \si{\hertz} and the current was measured with a SR830 lock-in amplifier. The chopper had a 50\% duty cycle and all reported laser power values were time-averaged to account for this. Low electromagnetic interference (EMI) piezo x-y stages were used to prevent interfering fields with the measurements. 

The chopper frequency-dependent photocurrent measurements were fit to the expression $I_f=I_0/\sqrt{1+(2\pi f \tau)^2}$, where $I_f$ is the measured current response, $f$ is the chopper frequency, $I_0$ is a fitting parameter, and $\tau$ is the time response. 

\subsection{Modeling}
To model absorption and electric field profiles in TaIrTe$_4$ we used Lumerical rigorous coupled wave analysis (RCWA) and finite-difference time-domain (FDTD) software packages.

Shockely-Ramo theory was used to model the local photocurrent response and combined with photonic simulations in COMSOL, detailed in Note~\ref{sec:SPCM-sim}.

For the 2D thermal modeling of the devices under IR illumination, we combined the transfer-matrix method to solve for the total optical absorption into TaIrTe$_4$ for $E\parallel a$ versus $E\parallel b$, accounting for the multilayer TaIrTe$_4$-SiO$_2$-Si heterostructure. We then solved the 2D heat equation at the 45\si{\degree} edge, treating the laser as a Gaussian heat source, the crystal edge as an insulating boundary, and accounting for the anisotropic material parameters in Note~\ref{sec:SPCM-sim}.

\section*{Acknowledgments}
M.G.B. thanks Dr. Mark Witinski, Elizabeth Gerrish, Dr. Shiekh Zia Uddin, and Dr. Sachin Vaidya for helpful experimental discussions. 
M.G.B. thanks Brady Cruse, Jacqueline Wang and Wyatt Vick for their assistance in electronics prototyping through the Undergraduate Research Opportunities Program at MIT (UROP).
X.J. thanks Dr. Yuxuan Wang for helpful discussions on photocurrent simulations. 
The authors thank the T.J. Rodgers RLE Laboratory for their helpful equipment and advice.

\section*{Author contributions}
M.G.B. fabricated devices, built the SPCM, conducted photocurrent, Raman spectroscopy and AFM measurements, and performed multiphysics modeling. 
X.J. conducted the photocurrent simulations and conducted Raman measurements. 
V.J.S.M. collected ellipsometric data. 
A.M. assisted in device fabrication and conducted XRD measurements.
T.N. grew the TaIrTe$_4$.
M.L. and S.V.B. supervised and guided the project.

\section*{Funding Sources}
M.G.B. was supported as a Draper Scholar by the Charles Stark Draper Laboratory, Inc. This work was supported in part by the MIT MISTI-Poland Seed Fund, and carried out in part through the use of MIT.nano facilities. The scanning photocurrent microscope setup was designed and built with support from the U.S. Department of Energy (Award DE-FG02-02ER45977). 

\bibliography{TaIrTe4-refs} % Produces the bibliography via BibTeX.
 
% \onecolumngrid

\newpage

%%%%%%%%%% Merge with supplemental materials %%%%%%%%%%

% Stop adding to table of contents
\addtocontents{toc}{\protect\setcounter{tocdepth}{-5}}

\clearpage
% \pagebreak
% \newpage
\onecolumngrid
\begin{center}
\textbf{\large Supplementary Information for ``Large Transverse Thermoelectric Effect in Type-II Weyl Semimetal TaIrTe$_4$ Engineered for Photodetection"\\}
\bigbreak
Blevins, Ji, Santamaria-Garcia, Mukherjee, Nguyen, Li, and Boriskina
\end{center}
%%%%%%%%%% Merge with supplemental materials %%%%%%%%%%
%%%%%%%%%% Prefix a "S" to all equations, figures, tables and reset the counter %%%%%%%%%%
\setcounter{equation}{0}
\setcounter{figure}{0}
\setcounter{table}{0}
\setcounter{page}{1}
\setcounter{section}{0}
\makeatletter
\renewcommand{\thesection}{S\arabic{section}}
\renewcommand{\theequation}{S\arabic{equation}}
\renewcommand{\thefigure}{S\arabic{figure}}
\renewcommand{\thetable}{S\arabic{table}}
% \renewcommand{\bibnumfmt}[1]{[S#1]}
% \renewcommand{\citenumfont}[1]{S#1}
%%%%%%%%%% Prefix a "S" to all equations, figures, tables and reset the counter %%%%%%%%%%
% Restart table of contents for SI only

\vspace{1.5em}
\noindent\rule{\linewidth}{0.4pt}
\vspace{0.3em}
\noindent{\large\textbf{Contents}}
\vspace{0.3em}
% \noindent\rule{\linewidth}{0.4pt}
\vspace{0.8em}

\noindent S1. I-V curves\dotfill \pageref{sec:IV-curve}\\[0.4em]
\noindent S2. Material thickness characterization via AFM \dotfill \pageref{sec:AFM}\\[0.4em]
\noindent S3. Dielectric function of $\text{TaIrTe}_4$\dotfill \pageref{sec:dielectric}\\[0.4em]
\noindent S4. More SPCM results of device A\dotfill \pageref{sec:more-devA}\\[0.4em]
\noindent S5. SPCM simulation\dotfill \pageref{sec:SPCM-sim}\\[0.2em]
\hspace*{1.8em} S5.A. Parameters for 3D thermal simulation of TaIrTe$_4$ flakes \dotfill \pageref{sec:parameters}\\[0.2em]
\hspace*{1.8em} S5.B. Modeling photo-thermoelectric effect in TaIrTe$_4$\dotfill \pageref{sec:model-PTE}\\[0.2em]
\hspace*{1.8em} S5.C. Shockley--Ramo theorem\dotfill \pageref{sec:shockley-ramo}\\[0.2em]
\hspace*{1.8em} S5.D. FEM simulation details\dotfill \pageref{sec:FEM}\\[0.4em]
\noindent S6. Tolerance of SPCM simulation to Seebeck coefficient and conductivity value variation\dotfill \pageref{sec:Seebeck-scan}\\[0.4em]
\noindent S7. Predicted PTE in thin TaIrTe$_4$ ($t=30$\,nm)\dotfill \pageref{sec:30nm-flake-sims}\\[0.4em]
\noindent S8. Evidence of lower thermal boundary conductance $G_{\text{TaIrTe}_4\text{-SiO}_2}$ on evaporated SiO$_2$ step\dotfill \pageref{sec:TBC-raman} \\ [0.4em]
\noindent S9. Raman spectra of TaIrTe$_4$ on SiO$_2$ step\dotfill \pageref{sec:raman-line-sio2-step}
\vspace{0.8em}
\noindent\rule{\linewidth}{0.4pt}

\section{I-V curves} \label{sec:IV-curve}

\begin{figure*}[ht]
    \includegraphics[width=1\textwidth]{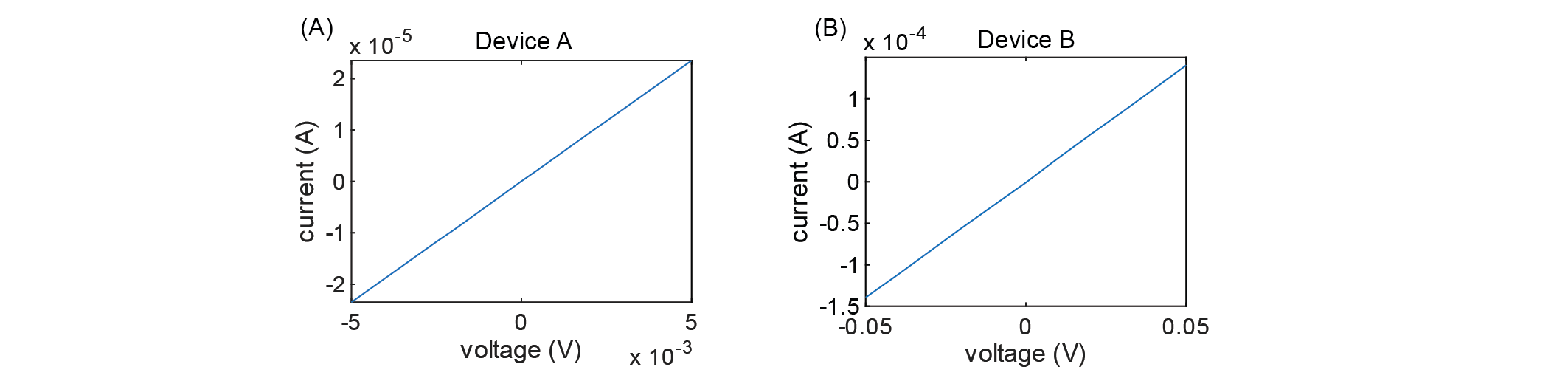}
    \caption{
    The measured I-V curves of device A ($R=213$~\si{\ohm}) and device B ($R=354$~\si{\ohm}).
    } 
    \label{fig:IV-curve}
\end{figure*}

The IV curves of devices A and B were collected using an Keithley 2400 Source Measure Unit (Figure~\ref{fig:IV-curve}). All devices have linear I-V curves, confirming good Ohmic contact.

\newpage
\clearpage

\section{Material thickness characterization via AFM} \label{sec:AFM}

\begin{figure*}[ht]
    \includegraphics[width=0.99\textwidth]{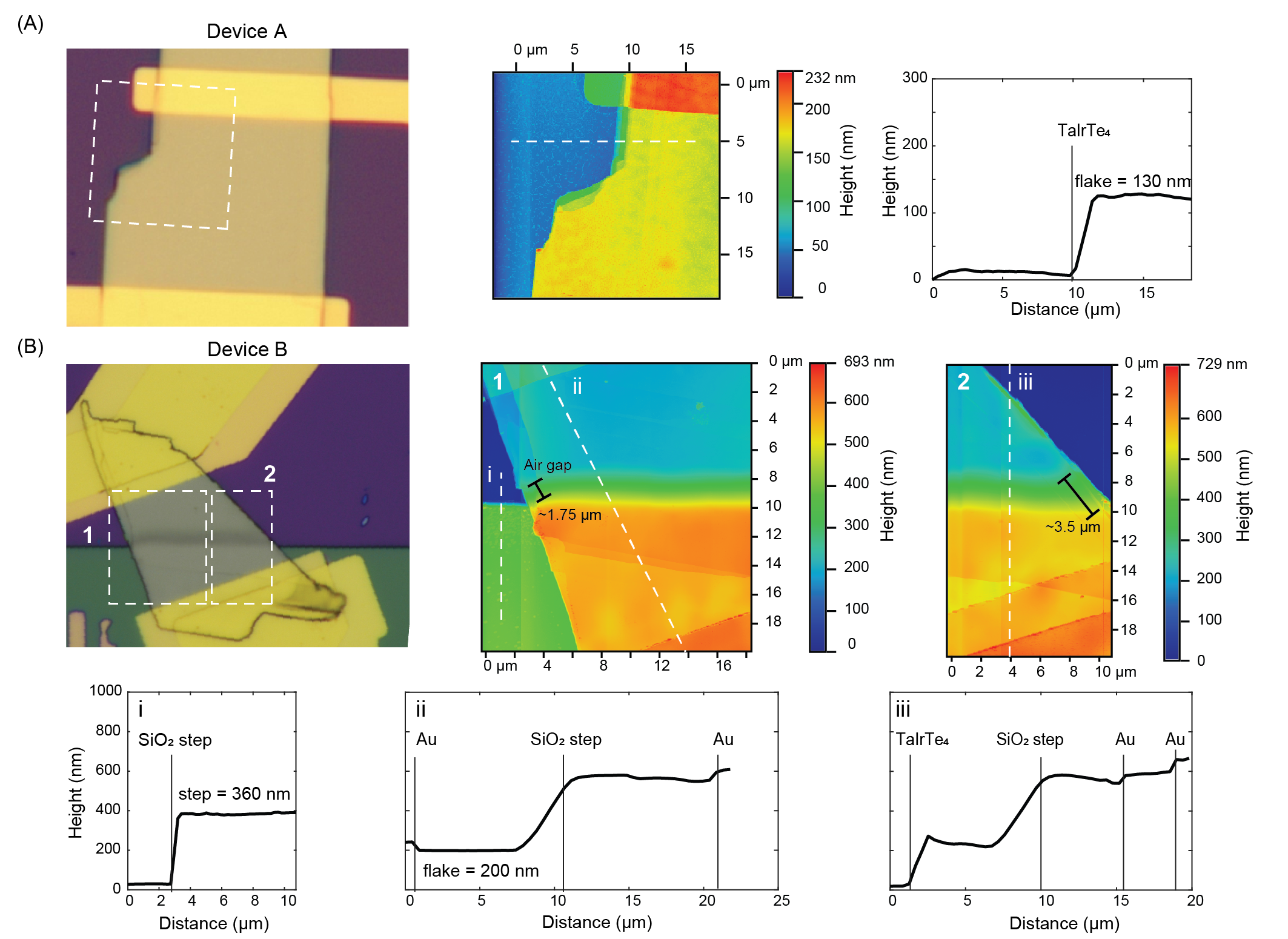}
    \caption{
    \textbf{AFM data of the TaIrTe$_4$ devices.} 
    (A) The height of the TaIrTe$_4$ in Device A was measured as 130 \si{\nano\meter}. (B) For device B, the height of the TaIrTe$_4$ was measured as 200 \si{\nano\meter} and the height of the SiO$_2$ step was measured as 360 \si{\nano\meter}. The air gap between the step and substrate is roughly 1.75~\si{\micro\meter} long on the left-side edge versus 3.5~\si{\micro\meter} on the right-side edge.
    } 
    \label{fig:afm}
\end{figure*}

The height of the TaIrTe$_4$ flakes were measured with an Asylum Instruments/Oxford Jupiter XR AFM (Table~\ref{tab:device-thicknesses}). The profiles of devices A and B are shown in Figure~\ref{fig:afm}. 

\begin{table}[ht]
\centering
\caption{Device descriptions and TaIrTe$_4$ flake thicknesses as measured with AFM.}
\label{tab:device-thicknesses}
\begin{tabular}{|c|c|l|c|}
\hline
\textbf{Device Name} & & \textbf{Description} & \textbf{Thickness (nm)} \\
\hline
Device A & & Currents at electrodes and at one edge (Figure~\ref{fig:635nm-PTE}A) & 130 \\
Device B & & Flake suspended on evaporated SiO$_2$-step (Figure~\ref{fig:sio2-step-635nm-results}A) & 200 \\
\hline
\end{tabular}
\end{table}

\newpage
\clearpage

\section{Dielectric function of $\text{TaIrTe}_4$} \label{sec:dielectric}

\begin{figure*}[ht]
    \includegraphics[width=0.5\textwidth]{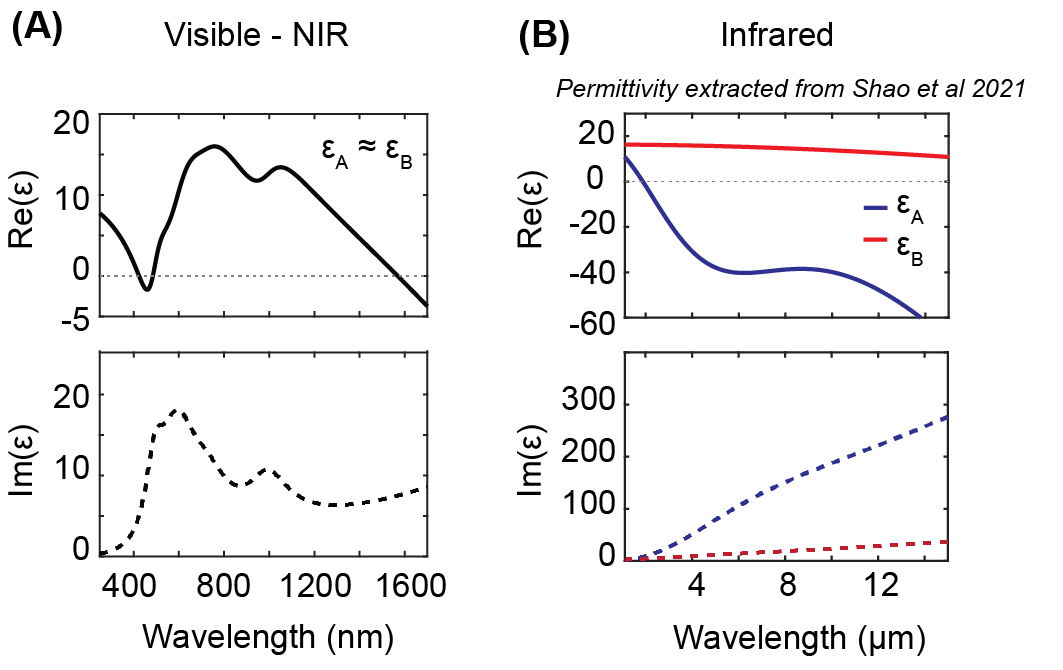}
    \caption{
    \textbf{Dielectric function of TaIrTe$_4$.} (A) The dielectric function was measured with the Accurion EP4 imaging ellipsometer across the visible and near infrared region. Here it is assumed that $\varepsilon_A \simeq \varepsilon_B$ for the 635~\si{\nano\meter} case (based on the results in Figure~\ref{fig:dev-A-635nm}B-D). (B) The anisotropic dielectric function $\varepsilon_A$ and $\varepsilon_B$ across the infrared are extracted from data in Ref.~\cite{ shao_NonlinearNanoelectrodynamicsWeyl_2021}.
    } 
    \label{fig:dielectric-func}
\end{figure*}

\begin{figure*}[ht]
    \includegraphics[width=0.99\textwidth]{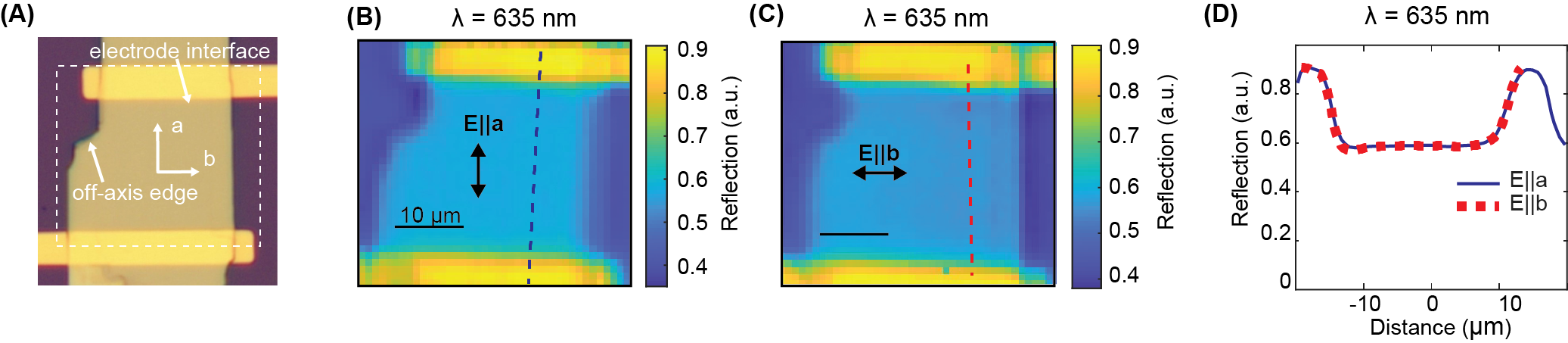}
    \caption{
    \textbf{Isotropic absorption at 635 nm in Device A.} 
    The SPCM reflection maps for (A) Device A were measured for both (B) $E \parallel a$ and (C) $E \parallel b$ (corresponding to the data in Figure~\ref{fig:635nm-PTE}E-F). (D) The reflection profiles across the device along the dashed lines in B-C show that reflection is equal for both polarizations, showing an isotropic absorption at this wavelength. 
    } 
    \label{fig:dev-A-635nm}
\end{figure*}

\noindent \textbf{Visible ---} The Accurion EP4 imaging ellipsometer was used to measure the permittivity, of TaIrTe$_4$ across the visible and near infrared (NIR) (Figure~\ref{fig:dielectric-func}A). The response was assumed isotropic for the visible range around $\lambda\sim$635~\si{\nano\meter} because in our reflection mapping we measured equal reflection from the TaIrTe$_4$ flakes independent of polarization angle at 635~\si{\nano\meter} (Figure~\ref{fig:dev-A-635nm}B-D).
\newline
\noindent \textbf{Infrared ---}
To evaluate the photonic response of TaIrTe$_4$ in the infrared we extracted the real and imaginary part of $\varepsilon_a$ and $\varepsilon_b$ from the data reported in Ref.~\cite{shao_NonlinearNanoelectrodynamicsWeyl_2021}.

\newpage
\clearpage

\section{More SPCM results of Device A} \label{sec:more-devA}

\begin{figure*}[ht]
    \includegraphics[width=0.99\textwidth]{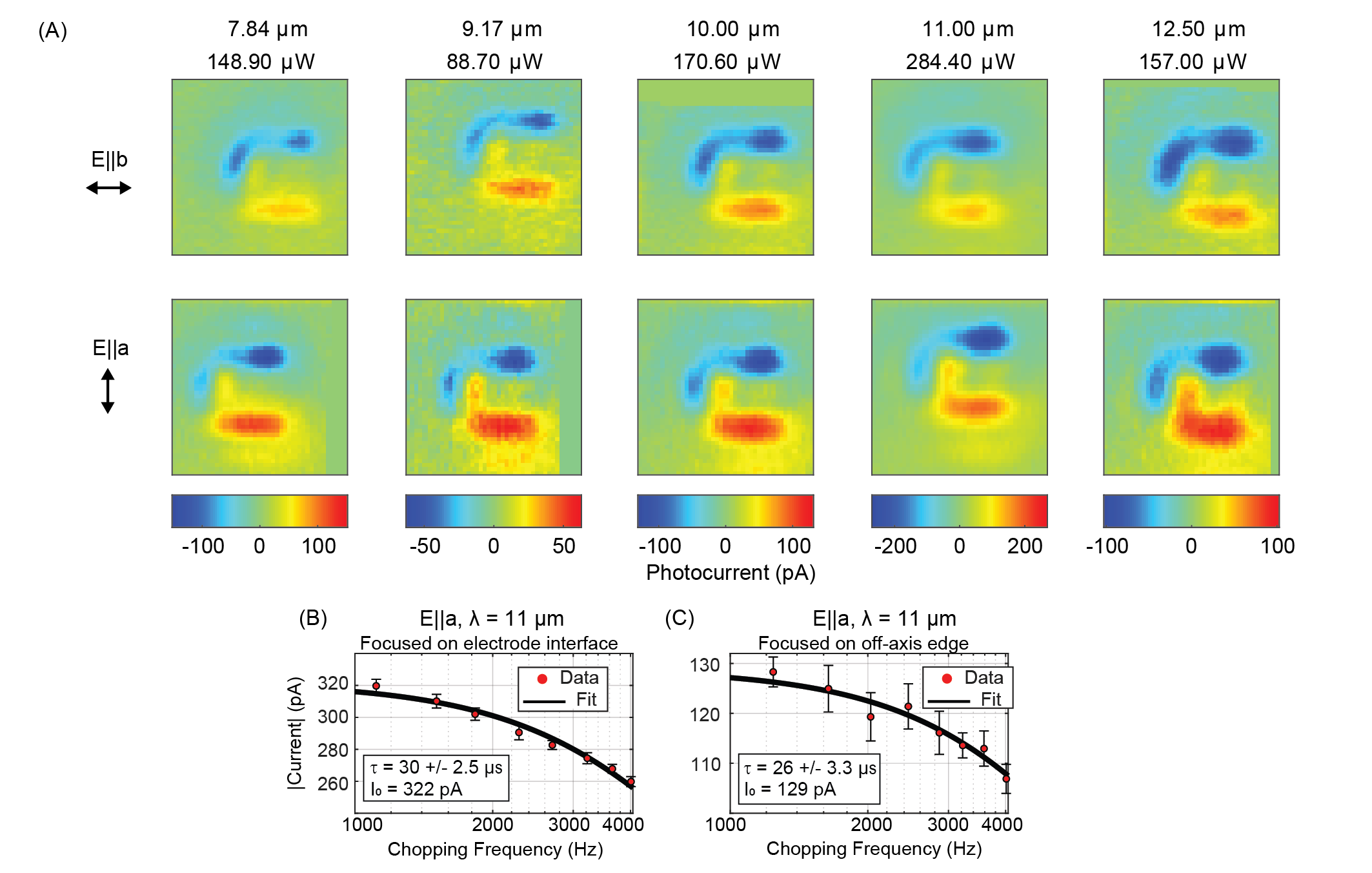}
    \caption{
    \textbf{LWIR SPCM results of Device A.} 
    (A) Full SPCM photocurrent maps were collected for $\lambda = 7.84, 9.17, 10, 11, 12.5$~\si{\micro\meter} laser illumination for $E\parallel a$ and $E\parallel b$. The data in Figure~\ref{fig:IR-results}J is extracted from the off-axis crystal edge region of these maps. Chopper frequency-dependent responsivity measurement were conducted at (B) the electrode interface and (C) at the device's off-axis crystal edge for the $E\parallel a$ case with $\lambda=11$ \si{\micro\meter}. Fitting the measured chopper frequency-dependent photocurrent trend to $I_f=I_0/\sqrt{1+(2\pi f \tau)^2}$, we found $\tau=30\pm 2.5$ \si{\micro\second} and $\tau=26\pm 3.3$ \si{\micro\second} for the interface and off-axis edge respectively. These large time response values are consistent with thermally driven currents. 
    } 
    \label{fig:dev-A-IR}
\end{figure*}

\subsection{Isotropic in-plane absorption at $\lambda=635$~\si{\nano\meter}}
The SPCM reflection maps of device A  for $E\parallel a$ and $E \parallel b$ at $\lambda=635$~\si{\nano\meter} show the same magnitude of reflection (Figure~\ref{fig:dev-A-635nm}A-D). 

\subsection{LWIR SPCM data}
The full SPCM photocurrent maps of device A under infrared illumination as well as the measured optical chopper frequency sweep are shown in Figure~\ref{fig:dev-A-IR}A-C. At each wavelength $I_b>I_a$ at the off-axis edge. However, due to the plasmonics response at the electrode interface for $E\perp \text{electrode}$ (which aligns with $E\parallel a$),  $I_a>I_b$ at the electrode interface.

\newpage
\clearpage

\section{SPCM simulation} \label{sec:SPCM-sim}

\subsection{Parameters for 3D thermal simulation of TaIrTe$_4$ flakes} \label{sec:parameters}

Parameters for the SPCM simulations are taken from our own measurements, vetted prior reports, or well-justified estimates. Table~\ref{tab:parameters} summarize the parameters used in the simulation. The Seebeck coefficients, $S_{a,b}$, in-plane thermal conductivity, $\kappa_{a,b}$, and electrical conductivity, $\sigma_{a,b}$, of TaIrTe$_4$ are from recently reported measurements~\cite{mutch_NTypeTransverseThermoelectrics_2022, zhu_AnisotropicHotCarrier_2025, zhang_RoomTemperatureFieldfree_2023}. 
The out of plane thermal conductivity, $\sigma_c$ , is estimated from a similar material \mbox{WTe$_2$}~\cite{liu_FirstprinciplesStudyLattice_2016}. The mass density, $\rho$, is computed from crystallography. For the visible range, permittivity $\varepsilon$ was measured by spectroscopic ellipsometry (Figure~\ref{fig:dielectric-func}A). In the mid-IR/IR, we use experimental permittivity values from Ref.~\cite{shao_NonlinearNanoelectrodynamicsWeyl_2021} (Figure~\ref{fig:dielectric-func}B). The constant-pressure heat capacity, $C_p$, is evaluated using the Dulong–Petit limit. Thermal boundary conductance, $G$, is estimated from similar material platforms. For the interface thermal conductance between TaIrTe$_4$ and evaporated SiO$_2$ step, we estimate that it is two orders of magnitude lower than that between TaIrTe$_4$ and the thermal SiO$_2$ layer on the silicon wafer~\cite{frausto-avila_ThermalBoundaryConductance_2022}. 

\begin{table}[!ht]
\centering
\begin{tabular}{l|r|c|c|c|c}
Description & Symbol & Value  & Unit & Ref \\\hline

Seebeck coefficient & $S_{a}$ & -6& \si{\micro\volt\per\kelvin} & \cite{mutch_NTypeTransverseThermoelectrics_2022} \\
~ & $S_{b} $ & 27 & \si{\micro\volt\per\kelvin} & \cite{mutch_NTypeTransverseThermoelectrics_2022} \\

Thermal conductivity & $\kappa_a$ & 14.4 & \SI{}{\watt\per\meter\per\kelvin}  & \cite{zhu_AnisotropicHotCarrier_2025} \\
~ & $\kappa_b$ & 3.8 & \SI{}{\watt\per\meter\per\kelvin}  & \cite{zhu_AnisotropicHotCarrier_2025}  \\
~ & $\kappa_c$ & 1.0 & \SI{}{\watt\per\meter\per\kelvin}  & \cite{liu_FirstprinciplesStudyLattice_2016} \\

Electrical conductivity & $\sigma_a$ &  $4.91 \times 10^5$ & \si{\per\ohm\per\meter}  & \cite{zhang_RoomTemperatureFieldfree_2023} \\
~ & $\sigma_b$ & $1.1\times10^5$ & \si{\per\ohm\per\meter}  & \cite{zhang_RoomTemperatureFieldfree_2023} \\

Density & $\rho$ & 9.45  & \SI{}{\gram\per\cubic\centi\meter} & ~ \\
Heat capacity & $C_p$ & 170  & \SI{}{\joule\per\kilo\gram\per\kelvin} & ~ \\
Interface thermal conductance & G$_{TaIrTe_4-therm~ SiO_2}$ & \num{7.37e6}, & \SI{}{\watt\per\square\meter\per\kelvin} & \cite{hunter_InterfacialThermalConductance_2020} \\
~ &  G$_{TaIrTe_4 - air}$ & \num{1}& \SI{}{\watt\per\square\meter\per\kelvin} & \cite{wang_VisualizationBulkEdge_2023} \\
~ & G$_{TaIrTe_4-evap~ SiO_2}$ & \num{7.37e4} & \SI{}{\watt\per\square\meter\per\kelvin} & \cite{kimling_ThermalConductanceInterfaces_2017} \\
\end{tabular}
\caption{\label{tab:parameters} TaIrTe$_4$ material parameters used for SPCM simulation. Values are for room temperature.}
\end{table}

\subsection{Modeling photo-thermoelectric effect in TaIrTe$_4$} \label{sec:model-PTE}
TaIrTe$_4$ exhibits strong in-plane anisotropy in its thermal, electrical, and optical response and we must take this into account when modeling the photocurrent response.
In scanning photocurrent microscopy (SPCM), a focused laser spot with Gaussian intensity heats the flake and generates in-plane temperature gradients. We model the laser profile at the surface as
\begin{equation}
\label{eq:laserI}
\begin{aligned}
I(x,y)=\frac{2P}{\pi w_0^2}\,\exp\!\left[-\frac{2(x^2+y^2)}{w_0^2}\right],
\end{aligned}
\end{equation}

\begin{equation}
\label{eq:laserQ}
\begin{aligned}
Q(x,y,z)=\frac{\eta P\,\beta\,e^{-\beta z}}{2\pi\sigma^2}\,\exp\!\left[-\frac{x^2+y^2}{2\sigma^2}\right],
\end{aligned}
\end{equation}
where $P$ is the incident power, $I(x,y)$ is incident laser intensity, $w_0=2\sigma$ is the beam radius, $\eta$ is the objective transmission, and $\beta$ is the absorption coefficient along $z$ (via Beer–Lambert law) that sets the volumetric heat source $Q(x,y,z)$. 
The 3D anisotropic heat equation is
\begin{equation}
\rho C_p\,\partial_t T-\nabla\!\cdot\!\big(\boldsymbol{\kappa}\,\nabla T\big)=Q(x,y,z),
\end{equation}
where $\rho$ is the density, $C_p$ is the heat capacity, and $\boldsymbol{\kappa}$ is the thermal conductivity matrix. At steady state we impose mixed  boundary conditions at the top and bottom interfaces to include the air/substrate heat conductances $G_{\rm int}$:
\begin{equation}
-\hat{\mathbf{n}}\cdot\boldsymbol{\kappa}\,\nabla T\big|_{\rm top} = G_{\rm top}\,\big(T-T_{\rm bath}\big),\qquad
-\hat{\mathbf{n}}\cdot\boldsymbol{\kappa}\,\nabla T\big|_{\rm bottom} = G_{\rm bottom}\,\big(T-T_{\rm bath}\big).
\end{equation}
This procedure matches prior thermal modeling used to analyze WTe$_2$/TaIrTe$_4$ devices under local heating and was shown to capture the measured temperature profiles under SPCM conditions \cite{wang_VisualizationBulkEdge_2023}. It can be further extended to perform 3D modeling so that heat transfer along z axis can be better simulated. 

In-plane second-order (shift) currents are symmetry-forbidden for normally incident light in the $ab$ plane for TaIrTe$_4$, so the dominant in-plane response under SPCM is photothermoelectric (PTE). Accounting for in-plane anisotropy, the current density is
\begin{equation}
\label{eq:apten}
\begin{aligned}
J_x(\mathbf{r})&=-\sigma_a\Big(\partial_x\Phi+S_a\,\partial_x T\Big),\\
J_y(\mathbf{r})&=-\sigma_b\Big(\partial_y\Phi+S_b\,\partial_y T\Big),
\end{aligned}
\end{equation}
where $\Phi$ is the electrochemical potential, $\sigma_{a,b}$ are the in-plane conductivities, and $S_{a,b}$ are the Seebeck coefficients along the crystal $a$/$b$ axes. Equation~\eqref{eq:apten} together with charge continuity $\nabla\!\cdot\!\mathbf{J}=0$ determines $\Phi(\mathbf{r})$ for a given $T(\mathbf{r})$.

\subsection{Shockley-Ramo theorem} \label{sec:shockley-ramo}

The current recorded at remote contacts is obtained by the Shockley–Ramo collection integral,
\begin{equation}
\label{eq:SR}
I_{\rm tot}=\iint \mathbf{J}_{\rm loc}(\mathbf{r})\cdot\nabla\psi(\mathbf{r})\,\mathrm{d}^2\mathbf{r},
\end{equation}
where $\psi(\mathbf{r})$ is the device-dependent weighting potential, and $\mathbf{J}_{\rm loc}=-\boldsymbol{\sigma}\,\mathbf{S}\nabla T$ is the local PTE source entering Eq.~\eqref{eq:SR}. Equation~\eqref{eq:SR} captures the long-range, geometry-sensitive collection of local currents routinely measured in SPCM and is agnostic to the microscopic generation mechanism. 

The weighting field $\mathbf{E}_w = -\nabla \psi_w$ is obtained by solving the Laplace equation
\begin{equation}
\label{eq:weightingfield}
\nabla^2 \psi(\mathbf{r}) = 0,
\end{equation}
under device-specific boundary conditions, with the collecting electrode set to unit potential
($\psi(\mathbf{r}) = 1$), all other electrodes grounded ($\psi(\mathbf{r}) = 0$), and insulating boundaries treated
with zero-flux conditions. Importantly, the weighting field depends only on the device
geometry and boundary conditions, and is independent of the microscopic details of charge
generation and transport.

Within this framework, spatially localized photocurrent sources---such as photothermoelectric currents induced by local temperature gradients---contribute to the measured electrode current according to their dot product with the weighting field. 

\subsection{FEM simulation details} \label{sec:FEM}

\begin{figure*}[ht]
    \includegraphics[width=0.5\textwidth]{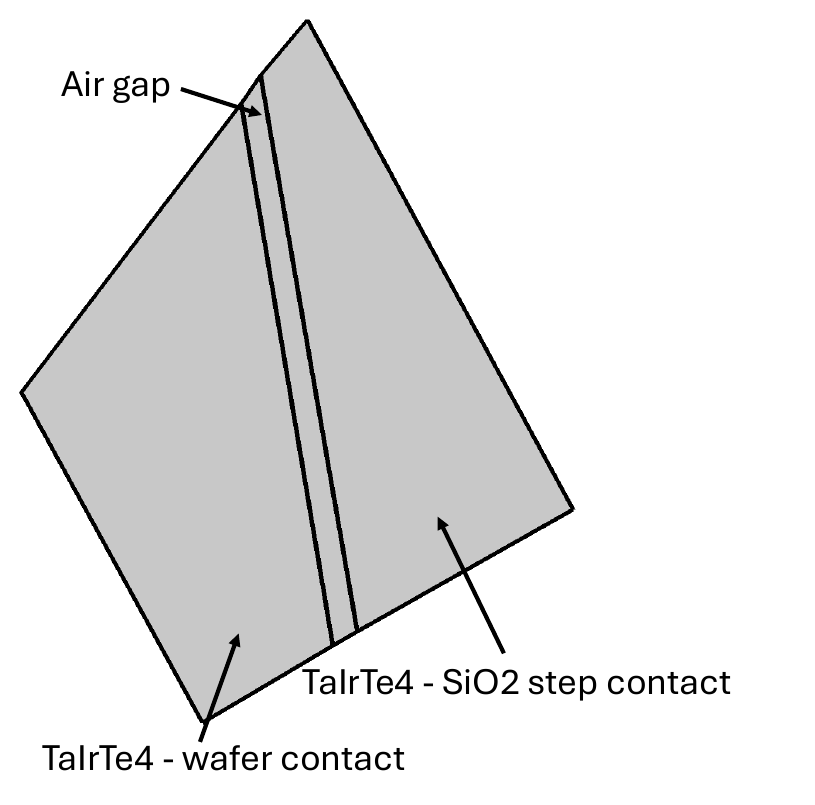}
    \caption{
    \textbf{COMSOL simulation setup for the thermally engineered device.} The thermal boundary conductance is defined for three regions of the thermally engineered device: the TaIrTe$_4$-substrate, TaIrTe$_4$-air, and TaIrTe$_4$-thermally evaporated SiO$_2$ interface.
    } 
    \label{fig:thermal_contacts}
\end{figure*}

Three-dimensional COMSOL Multiphysics simulations are performed to numerically calculate the scanning photocurrent microscopy (SPCM) response. The Radiative Beam in Absorbing Media interface is used to compute the volumetric heat source generated by the incident laser, accounting for optical absorption through the spatial attenuation of the beam. The laser beam is modeled with a Gaussian profile (Equation \ref{eq:laserI}) in the x-y plane and an exponential decay along the z direction (Equation \ref{eq:laserQ}). The linear absorbtion coefficient is calculated from the imaginary part of refractive index under different wavelength. All absorbed optical power is assumed to be converted into heat, neglecting carrier energy storage and radiative recombination, which is a valid approximation for photothermoelectric transport. Experimental parameters of the visible and infrared lasers, including incident power and beam size, are used as simulation inputs. The resulting temperature distribution is calculated using the Heat Transfer in Solids module. Heat flux boundary conditions are applied at the interfaces, defined by different thermal boundary conductance corresponding to different contacts. The boundary condition for the thermally engineered device is shown is Figure~\ref{fig:thermal_contacts}(c).

The weighting field in Equation \ref{eq:weightingfield} is obtained by solving this Laplace equation in COMSOL under device-specific boundary conditions. The sample–air boundaries are treated with zero-flux (Neumann) conditions, while the sample–electrode boundaries are assigned Dirichlet boundary conditions, with one electrode set to unit potential (1) and the other grounded (0).
The photocurrent distribution is then obtained by multiplying the local photocurrent vector field by the weighting field and performing a volumetric integration over the entire device (Equation \ref{eq:SR}).

\newpage
\clearpage

\section{Tolerance of SPCM simulation to Seebeck coefficient and conductivity value variation} \label{sec:Seebeck-scan}

Here we show that our SPCM current simulations and the main conclusion of our manuscript remain valid when parameter values are slightly varied. 

\subsection*{Justification of anisotropic Seebeck coefficients in-plane}

Because TaIrTe$_4$ is a type-II Weyl semimetal, its Weyl points occur at the band crossing point between electron and hole pockets~\cite{soluyanov_TypeIIWeylSemimetals_2015,koepernik_TaIrTe4TernaryTypeII_2016}. Notably, the electron and hole pockets have different dimensionality in the Fermi surface for TaIrTe$_4$~\cite{koepernik_TaIrTe4TernaryTypeII_2016}. The coexistence of these hole and electron Fermi surfaces with different dimensionality is crucial for the realization of a large transverse thermoelectric effect in the crystal~\cite{manako_LargeTransverseThermoelectric_2024}.

While the anisotropic Seebeck coefficients are taken from the preprint Ref.~\cite{mutch_NTypeTransverseThermoelectrics_2022}, the coexistence of the electron/hole pockets with different dimensionality in TaIrTe$_4$ is presented in several peer-reviewed articles~\cite{khim_MagnetotransportHaasvanAlphen_2016, koepernik_TaIrTe4TernaryTypeII_2016} and justifies the conclusion of anisotropic thermal response in TaIrTe$_4$. 

\subsection*{Sensitivity of SPCM maps to variation of the Seebeck coefficients and conductivity values}

Here, we show the sensitivity of the simulated SPCM current pattern to variations of the Seebeck coefficients $S_a=-6$~\si{\micro\volt\per\kelvin} and  $S_b=27$~\si{\micro\volt\per\kelvin} reported in the preprint Ref.~\cite{mutch_NTypeTransverseThermoelectrics_2022} and the conductivity values $\sigma_a= 4.9 \times 10^5$~\si{\per\ohm\per\meter} and $\sigma_b= 1.1 \times 10^5$~\si{\per\ohm\per\meter} reported in Ref.~\cite{zhang_RoomTemperatureFieldfree_2023}. 
We calculated the SPCM photocurrent maps for Device A with a 2-D Shockley-Ramo simulation, accounting for $\Delta T(x,y)$ induced from a laser with spot size 1~\si{\micro\meter}. We show that the simulated current pattern is robust against slight variations of the Seebeck coefficients and conductivity values.

\begin{figure*}[ht]
    \includegraphics[width=1\textwidth]{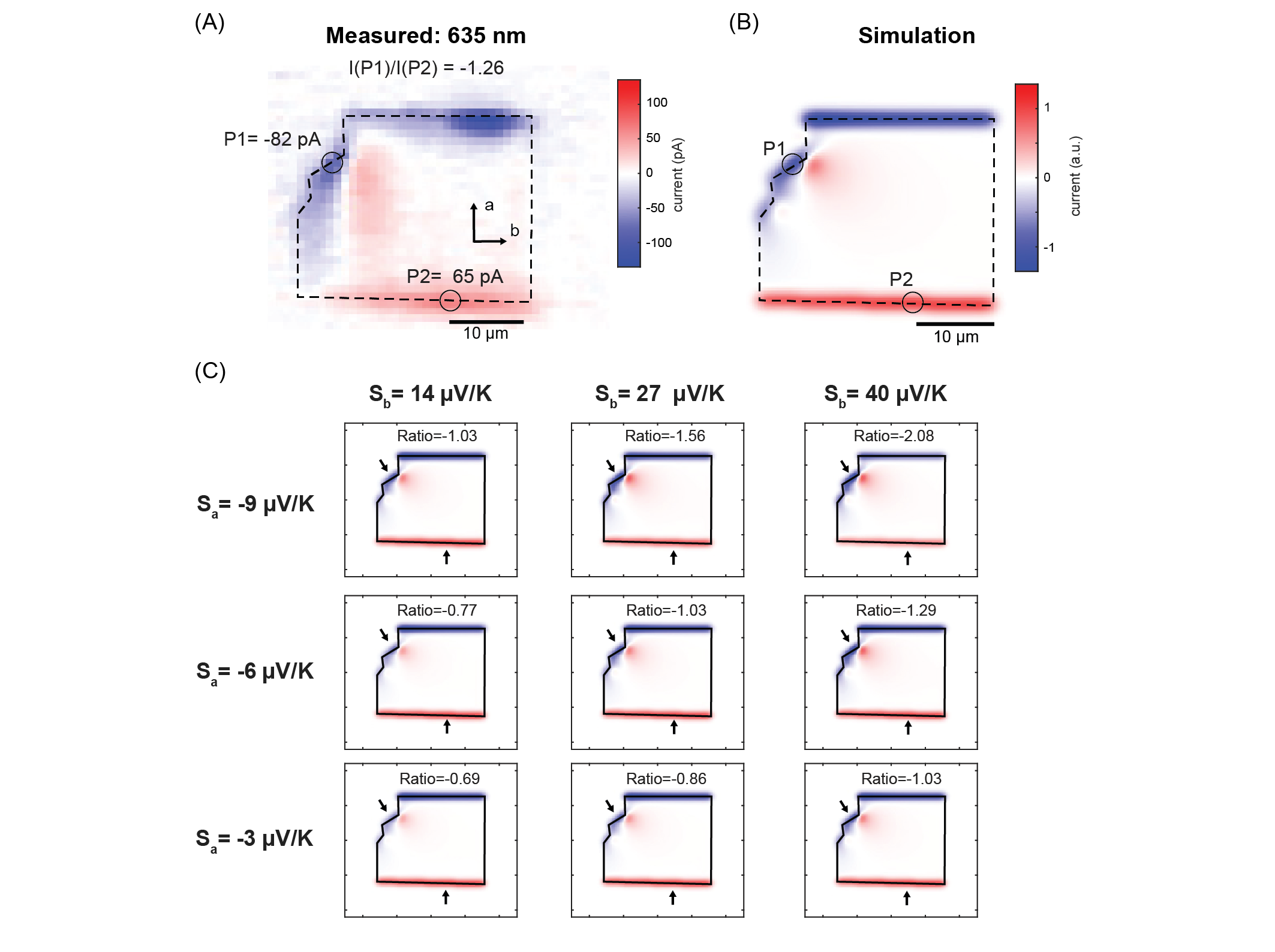}
    \caption{
    \textbf{SPCM simulation variation with changing Seebeck coefficients.} (A) The measured SPCM current map ($\lambda=635$~\si{\nano\meter}) of Device A has a current ratio at the off-axis edge (point $P1$) over that at the bottom electrode (point $P2$) of approximately $I(P1)/I(P2)=-1.26$.
    (B) The 2-D Shockley-Ramo simulation of Device A, using the Seebeck coefficients $S_a=-6$~\si{\micro\volt\per\kelvin} and $S_b=27$~\si{\micro\volt\per\kelvin} reported in the preprint Ref.~\cite{mutch_NTypeTransverseThermoelectrics_2022} and the conductivity values $\sigma_a= 4.9 \times 10^5$~\si{\per\ohm\per\meter} and $\sigma_b= 1.1 \times 10^5$~\si{\per\ohm\per\meter} reported in Ref.~\cite{zhang_RoomTemperatureFieldfree_2023}. The simulation has a ratio of approximately $I(P1)/I(P2)=-1.03$, in reasonable agreement with the experimental result.
    (C) While maintaining the conductivity values, the SPCM current pattern was simulated across the range of $S_a=-6\pm50\%$~\si{\micro\volt\per\kelvin} and $S_b=27\pm50\%$~\si{\micro\volt\per\kelvin}. This alters the predicted $I(P1)/I(P2)$ ratio while maintaining the same qualitative photocurrent pattern.}
    \label{fig:seebeck_maps}
\end{figure*}

\begin{figure*}[ht]
    \includegraphics[width=1\textwidth]{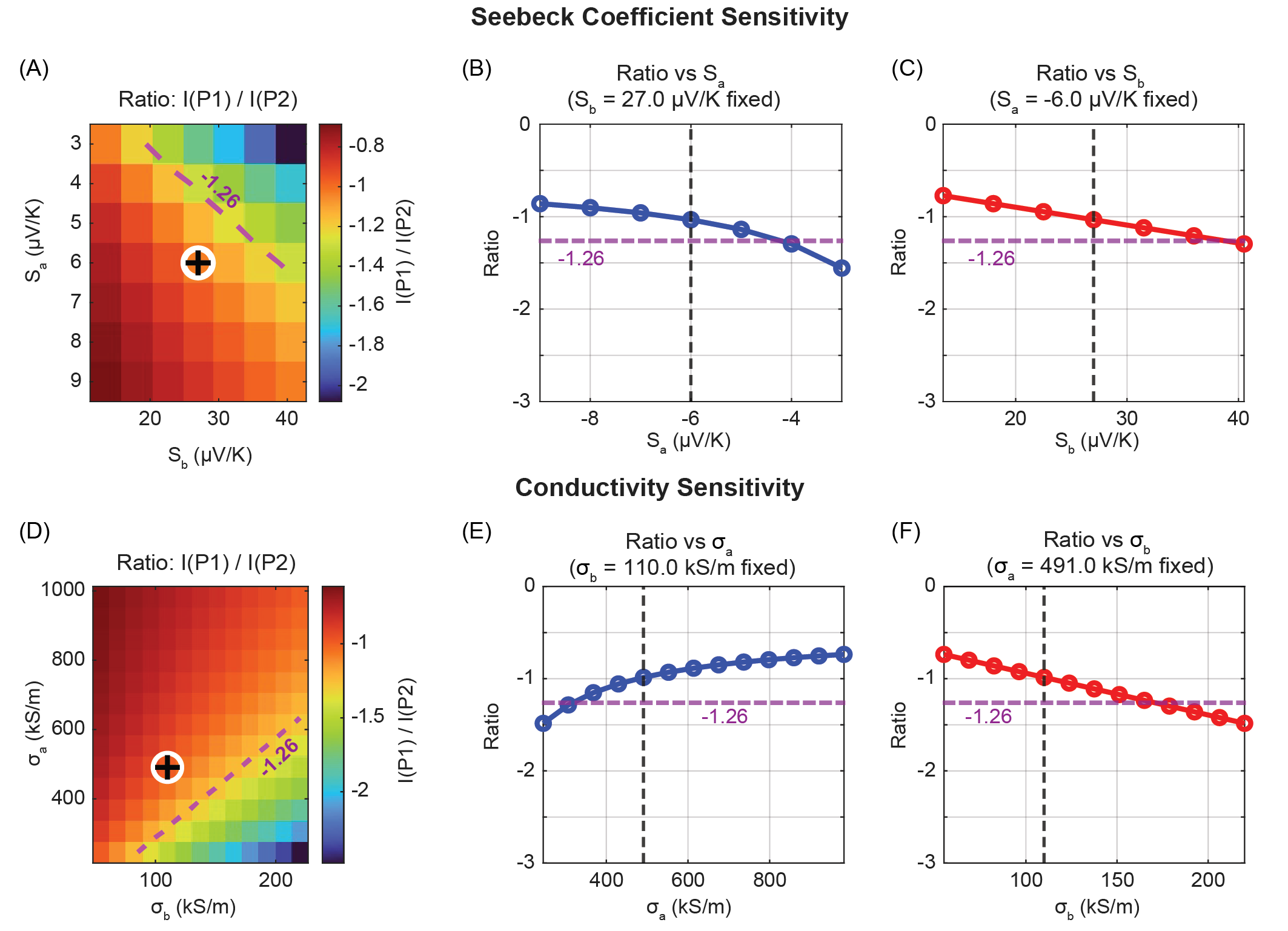}
    \caption{
    \textbf{Sensitivity of SPCM simulation to Seebeck coefficients and conductivity values.} 
    (A) The value of $I(P1)/I(P2)$ as $S_a$ and $S_b$ are varied and $\sigma_a$ and $\sigma_b$ are held constant. The pink dashed contour line shows the $S_a$ and $S_b$ values for which the simulation matches the experimental value of $I(P1)/I(P2)=-1.26$. The white circle and black cross show the value for the simulation parameter used in the manuscript.
    (B) While $S_b$ is fixed, the simulated ratio varies slightly as $S_a$ decreases and approaches an asymptote as $S_a\rightarrow0$. The dashed magenta line indicates the experimentally measured ratio of $-1.26$ and the dashed black vertical line marks the base value $S_a = -6$ \si{\micro\volt\per\kelvin}.
    (C) When $S_a$ is fixed, the ratio is largely insensitive to variations in $S_b$ across the swept range, suggesting that the ratio at these points is largely governed by $S_a$ rather than $S_b$. The dashed black vertical line marks the base value $S_b = 27$ \si{\micro\volt\per\kelvin}.
    (D) The value of $I(P1)/I(P2)$ as $\sigma_a$ and $\sigma_b$ are varied and $S_a$ and $S_b$ are held constant. The pink dashed contour line shows the $\sigma_a$ and $\sigma_b$ values for which the simulation matches the experimental value of $I(P1)/I(P2)=-1.26$.
    (E) When $\sigma_b$ is fixed, the ratio increases monotonically with $\sigma_a$.
    (F) When $\sigma_a$ is fixed, the ratio decreases monotonically with $\sigma_b$.}
    \label{fig:seebeck_cond_sensitivity}
\end{figure*}

We sought to determine the variation of the ratio of current at the off-axis edge (point $P1$) to that at the electrode interface (point $P2$), $I(P1)/I(P2)$, (shown in Figure~\ref{fig:seebeck_maps}A-B) over a range of [$S_a$, $S_b$] and [$\sigma_a$, $\sigma_b$] values. We evaluated the current ratio of these two points because this is the metric that is tied to the transverse photo-thermoelectric effect (the ratio goes to zero for an isotropic material). Further, ratios are appropriate for this evaluation because we seek to capture the current pattern and not the absolute magnitude of current in our modeling. 
The results of this study are shown in Figures~\ref{fig:seebeck_maps} and \ref{fig:seebeck_cond_sensitivity}.
\newline \newline
\noindent \textbf{{Seebeck coefficient sensitivity:}}
\begin{itemize}
    \item We calculated the SPCM photocurrent maps over the range $S_a=-6\pm50\%$~\si{\micro\volt\per\kelvin} and $S_b=27\pm50\%$~\si{\micro\volt\per\kelvin}, while maintaining $\sigma_a= 4.9 \times 10^5$~\si{\per\ohm\per\meter} and $\sigma_b= 1.1 \times 10^5$~\si{\per\ohm\per\meter}. 
    We selected this Seebeck coefficient range because it extends beyond a reasonable uncertainty bound around the values measured in Ref.~\cite{mutch_NTypeTransverseThermoelectrics_2022}.
    \item Experimentally, the ratio is approximately $I(P1)/I(P2)=-1.26$. Using the room temperature data from Ref.~\cite{mutch_NTypeTransverseThermoelectrics_2022}, stated above, our simulation results give $I(P1)/I(P2)=-1.03$, which is in reasonable agreement with our experiment (Figure~\ref{fig:seebeck_maps}A-B).
    \item As S$_a$ and S$_b$ are varied, the ratio $I(P1)/I(P2)$ is modulated as shown by the grid of SPCM simulations in Figure~\ref{fig:seebeck_maps}C and the results in Figure~\ref{fig:seebeck_cond_sensitivity}A-C. As $S_a\rightarrow0$, the current at the electrode will go to zero and this drives a large $|I(P1)/I(P2)|$ (top of the Figure~\ref{fig:seebeck_cond_sensitivity}A heatmap).
    \item The ratio of $I(P1)/I(P2)=-1.26$ is preserved over the pink contour line shown in Figure~\ref{fig:seebeck_cond_sensitivity}A. 
    This means that the same quantitative PTE current pattern is captured even if the exact anisotropy of the S$_a$ and S$_b$ values and their difference is varied.
\end{itemize}

\noindent \textbf{Conductivity sensitivity:}

\begin{itemize}
    \item Here, we calculated the SPCM photocurrent maps over the multiplicative range of $0.5 - 2.0 \times$ $[\sigma_a= 4.9 \times 10^5$~\si{\per\ohm\per\meter}, $\sigma_b= 1.1 \times 10^5$~\si{\per\ohm\per\meter}$]$ while maintaining $S_a=-6$ \si{\micro\volt\per\kelvin} and $S_b=27$ \si{\micro\volt\per\kelvin} (Figure~\ref{fig:seebeck_cond_sensitivity}D-F). We selected this range because conductivities can span orders of magnitude.
    \item Figure~\ref{fig:seebeck_cond_sensitivity}D shows that the ratio of $I(P1)/I(P2)=-1.26$ is preserved over the drawn pink contour line. This means that the same quantitative PTE current pattern is captured even if the exact conductivity values are varied by about $~\sim0.5-2\times$ for $\sigma_a$ and $\sim0.5-1.0\times$ for $\sigma_b$.
\end{itemize}

These results show that our theoretical simulations and interpretation of PTE driven photocurrent are robust against slight variations in $S_a$, $S_b$, $\sigma_a$, and $\sigma_b$.

\newpage
\clearpage

\section{Predicted PTE in thin $\text{TaIrTe}_4$ ($t= 30 \text{nm}$)} \label{sec:30nm-flake-sims}

\begin{figure*}[ht]
    \includegraphics[width=1\textwidth]{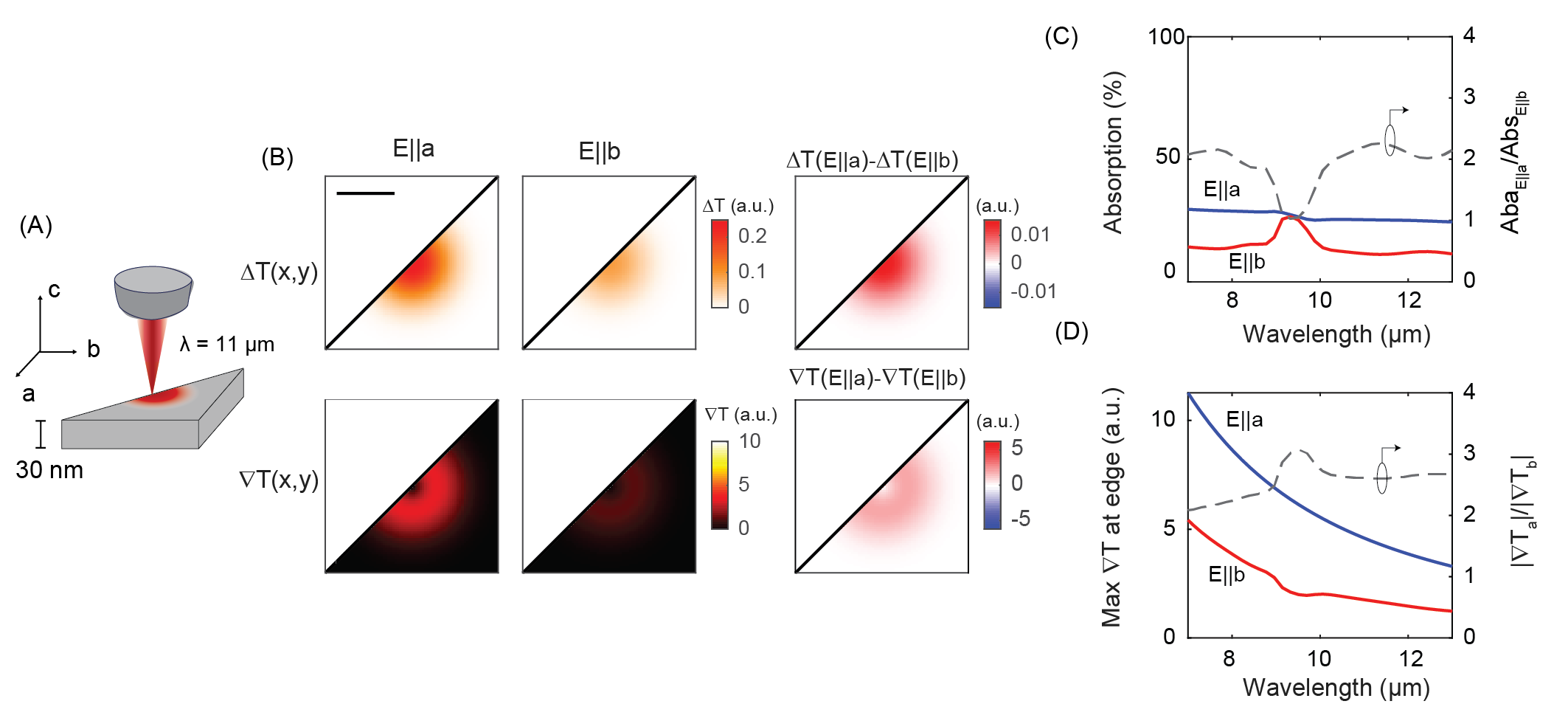}
    \caption{\textbf{2D thermal model of laser heating in 30 nm-thick TaIrTe$_4$.}
    (A) For a TaIrTe$_4$ flake of 30 nm thickness, (B) the local temperature change and temperature gradient driven by $\lambda=11$~\si{\micro\meter} illumination is larger for $E\parallel a$ than $E\parallel b$, the opposite trend as that predicted for thicker TaIrTe$_4$ (as in Figure~\ref{fig:IR-results}). (C) Optical absorption as calculated with the TMM and (D) maximum temperature gradient at the 45\si{\degree} crystal edge are larger for $E\parallel a$ in the LWIR. This predicts larger PTE currents for $E\parallel a$ and is consistent with the IR SPCM measurements of the thin flakes in Refs.~\cite{ma_NonlinearPhotoresponseTypeII_2019, lai_BroadbandAnisotropicPhotoresponse_2018,deng_TuningDecayLength_2025}.}
    \label{fig:IR-sim-30nm}
\end{figure*}

\newpage
\clearpage

\section{Evidence of lower thermal boundary conductance $\text{G}_{\text{TaIrTe}_4-\text{SiO}_2}$ on evaporated SiO$_2$-step} \label{sec:TBC-raman}

\begin{figure*}[ht]
    \includegraphics[width=0.5\textwidth]{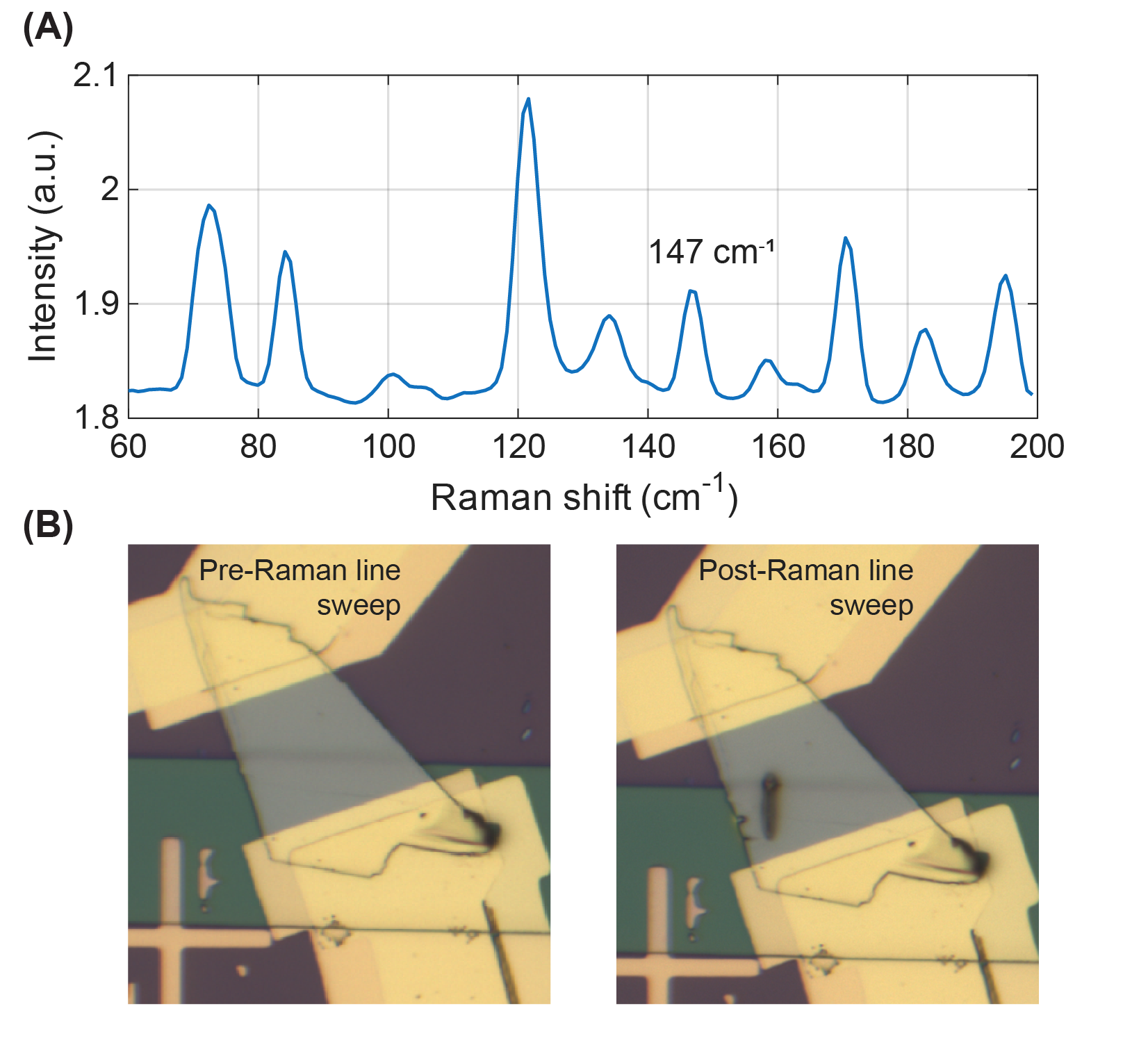}
    \caption{
    (A) The Raman spectrum of TaIrTe$_4$. (B) The SiO$_2$ step-suspended flake before (left) and after (right) taking a line of Raman data from the SiO$_2$ step side to the substrate suspended side.
    } 
    \label{fig:raman-spectrum-burn}
\end{figure*}

As evidence of the change in thermal boundary conductance $\text{G}_{\text{TaIrTe}_4-\text{SiO}_2}$ between the evaporated SiO$_2$ step side and the thermally grown SiO$_2$ side of the flake in device B shown in Figure~\ref{fig:sio2-step-635nm-results}A, we show the result of taking a Raman line sweep across the device (Figure~\ref{fig:raman-spectrum-burn}B). We optimized the Raman laser power (4~\si{\milli\watt}) and accumulation time (3~\si{\second}) on the substrate-suspended side and took several 73-point sweep measurements on that side. We then used the same laser parameters to conduct a line sweep with 500~\si{\nano\meter} steps across the full device. This resulted in the burning shown in Figure~\ref{fig:raman-spectrum-burn}(B). The burning is isolated to the evaporated SiO$_2$-step side. This suggests that the local heating from the laser was much higher on that side than the substrate side. We interpret this to mean that the thermal boundary conductance of the evaporated SiO$_2$ is much lower than that of the thermally grown SiO$_2$ of the substrate.

Note that the burning during the Raman line sweep prevented us from profiling the exact strain in this device. We thus profiled another, thicker TaIrTe$_4$ device which withstood Raman in Sec.~\ref{sec:TBC-raman}.

\newpage
\clearpage

\section{Raman Spectra of $\text{TaIrTe}_4$ on $\text{SiO}_2$ step} \label{sec:raman-line-sio2-step}

\begin{figure*}[ht]
    \includegraphics[width=1\textwidth]{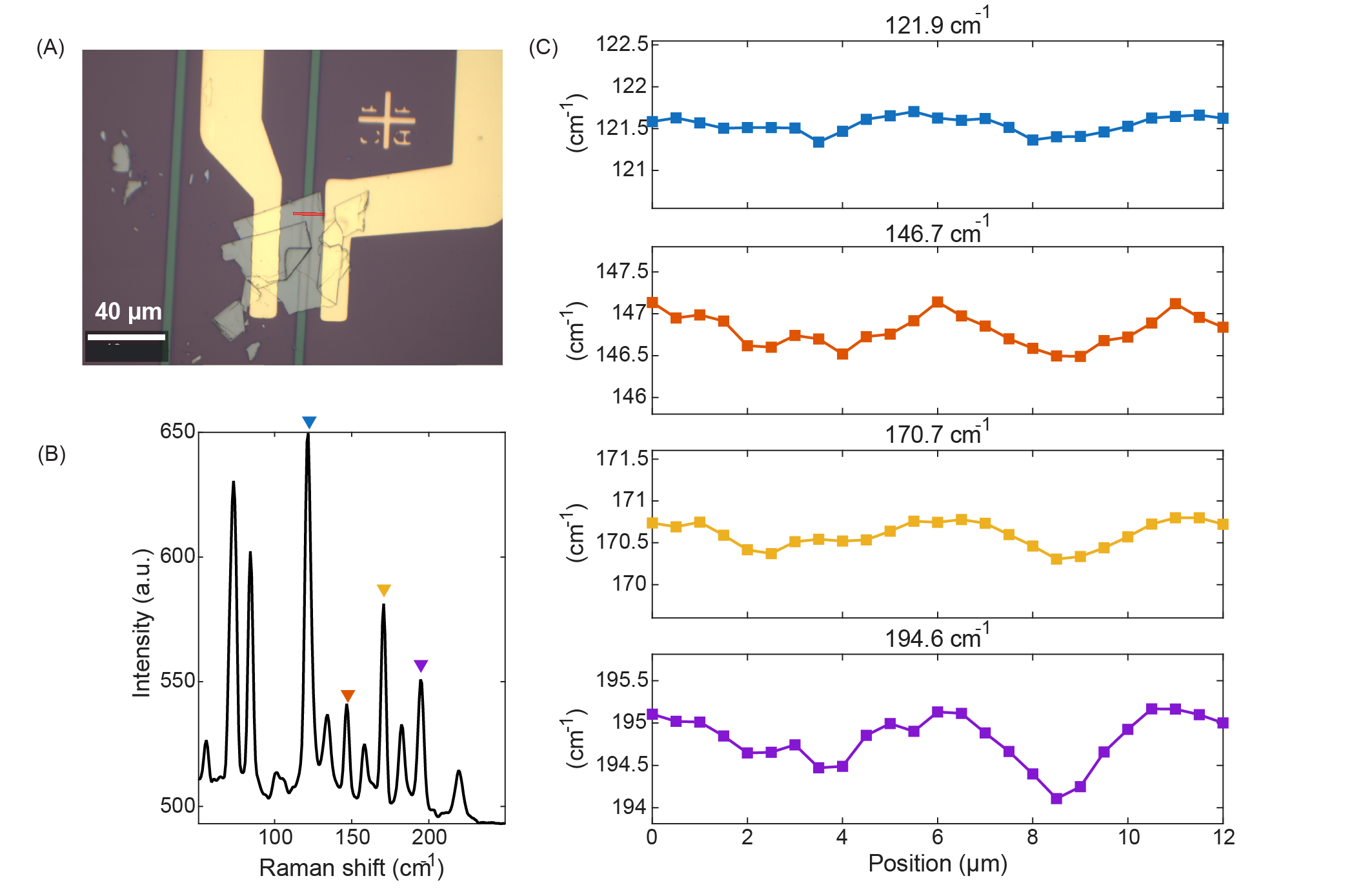}
    \caption{(A) A TaIrTe$_4$ flake is conformed to a 360~\si{\nano\meter} step. (B) The Raman spectra was collected along the red line to probe the crystal lattice on and off the step. (C) The Raman peak positions at 121.9, 146.7, 170.7 and 194.6~\si{\per\centi\meter} were monitored across the line sweep and show slight shifts at the step-flake transitions.
    } 
    \label{fig:raman-line}
\end{figure*}

Raman spectra were collected along a line scan across a TaIrTe$_4$ flake suspended over a 360~\si{\nano\meter} SiO$_2$ step to investigate potential strain in the crystal lattice from conformation to the step geometry. Raman spectra were collected every 500~\si{\nano\meter} across the 12~\si{\micro\meter} red line shown in Figure~\ref{fig:raman-line}A. 
We monitored the peak position of the four Raman peaks labeled in Figure~\ref{fig:raman-line}B. Each peak is modulated by $\sim0.25$~\si{\per\centi\meter} across the SiO$_2$ step. 
In terms of lattice strain, this magnitude of peak position shift is associated with $\sim1-4\%$ strain in a sister Weyl semimetal of the same symmetry group WTe$_2$ ~\cite{yang_AnomalousEnhancementNernst_2023}. However, any possible strain appears confined to the immediate transition regions (positions $\sim4$ and $\sim8$~\si{\micro\meter} in Figure~\ref{fig:raman-line}C) at the step edges, whereas our analysis focuses on comparing the response within the flat on-step and off-step regions. 

\end{document}